\documentclass{article}
\usepackage{emulateapj}
\usepackage{graphicx}

\slugcomment{To appear in ApJ}

\begin{document}


\title{Spectroscopic Signatures of Convection in the Spectrum of Procyon. 
Fundamental Parameters and Iron Abundance.}

\author{Carlos Allende Prieto}
\affil{McDonald Observatory and Department of Astronomy \\ University of Texas
  \\ RLM 15.308, Austin, TX 78712-1083, \\ USA}

\author{Martin Asplund}
\affil{Uppsala Astronomical Observatory \\ Box 515, S-751 20 
Uppsala \\ Sweden}

\author{Ram\'on J. Garc\'{\i}a L\'opez\altaffilmark{1}}
\altaffiltext{1}{Departamento de Astrof\'{\i}sica, Universidad de La Laguna, 
E-38206 La Laguna, Spain}
\affil{Instituto de Astrof\'\i sica de Canarias \\ E-38200, La Laguna,
Tenerife,  \\  Spain}

\and

\author{David L. Lambert}
\affil{McDonald Observatory and Department of Astronomy \\ University of Texas
\\ RLM 15.308, Austin, TX 78712-1083, \\ USA}

\begin{abstract}

We have observed the spectrum of Procyon A (F5IV) 
from 4559 to 5780 \AA\ with a $S/N$ of $\sim   10^3$ and a 
resolving power of $2 \times 10^5$. We
have measured the line bisectors and relative line shifts 
of a large number of Fe I and Fe II lines, comparing them 
to those found in the Sun. A  three-dimensional  (3D) 
hydrodynamical model atmosphere has been computed and is tested against
 observations. The model reproduces in detail most of the
  features observed, although we identify some room for improvement. 
At all levels, the comparison of the 3D time-dependent calculations 
with the observed spectral lines shows a much better
agreement than for classical homogeneous models, making
it possible to refine previous estimates of the iron 
abundance, the projected rotational velocity, 
the limb-darkening, and 
 the systemic velocity of the Procyon binary system. 
 The difference 
 between the iron abundance determined with the 3D model and its 
 1D counterpart is $\lesssim 0.05$ dex. We find consistency between
 the iron abundance derived from Fe I and Fe II lines, 
suggesting that departures from LTE in 
 the formation of the studied lines are relatively small.  The scatter 
 in the iron abundance 
 determined from different lines still exceeds the expectations from the 
 uncertainties in the atomic data, pointing out that one or more components  
 in the modeling can be refined further.
\end{abstract}


\section{Introduction}

Widely available grids of  model atmospheres  are based on 
a  classical set of  assumptions that have remained largely 
unchanged for many years. In brief, the classical assumptions are
that  the atmosphere consists of plane-parallel homogeneous layers in
hydrostatic and local thermodynamic equilibrium through which a constant
(radiative plus convective) flux is transmitted.
This is not to say that model atmospheres have not
been progressively refined; one notes, for example, the increasing
completeness with which line blanketing is included as an opacity source.
 The geometric assumption  
of plane parallel homogeneous layers  
has been replaced by the adoption of spherically
homogeneous layers for construction of model atmospheres
of supergiants and giants. More recently, the   
assumption of local thermodynamic equilibrium (LTE) has been replaced
to a limited extent by that of non-LTE, largely for models built 
to represent individual stars.
  Another classical assumption
-- hydrostatic equilibrium --  has  more tenaciously resisted refinement.
This paper discusses models that dispense with this assumption (and the
geometric assumption of homogeneous layers) by simulating the surface convection as a 
time-dependent hydrodynamic phenomenon.

The models representing  atmospheric convection are a development of
three-dimensional hydrodynamical
 models that realistically represent  solar granulation 
(Stein \& Nordlund 1998; Asplund et al. 2000a). 
Previous extensions of the
models to  stars have been discussed by
 Dravins \& Nordlund (1990) (Procyon A, $\alpha$ Cen A and B, and $\beta$ Hyi) 
and Asplund et al. (1999) (metal-poor stars  HD\,84937 and HD\,140283).
In this paper, we confront the models with an extensive set of line
profiles of Procyon A obtained at very high-resolution and high
signal-to-noise ratio. Our tests are similar to those discussed by
Dravins \& Nordlund but provide a much stiffer challenge of the models.

Crudely,  the 3D hydrodynamical model
atmosphere comprises  areas of hot rising gas (granules) and  cool sinking gas
(inter-granular lanes). Relative to a line
profile from a classical atmosphere,
the lines are broader, shifted,  asymmetric, and of a different
equivalent width. Derivation of fundamental parameters, 
abundances, and other quantities
extracted from observed spectra has not been attempted using
3D stellar model atmospheres with a few exceptions
(e.g. Bruls \& Rutten 1992; Atroshchenko \& Gadun 1994; 
Kiselman \& Nordlund 1995; 
Shchukina \& Trujillo Bueno 2001). This may be a  serious omission, because 
modeling the effect of the convective
motions and temperature inhomogeneities on the equivalent widths through the
 parameters known as micro- and macroturbulence may represent an important
source of systematic errors.  To the extent that equivalent widths  
are used to
derive defining fundamental parameters,  classical and 3D hydrodynamical models
will yield different results.
For the solar case, however, Allende Prieto et al. (2001) 
have shown that the
abundances derived using a numerical inversion based on a 1D model 
atmosphere are accurate to within 0.04 dex. Naturally, this will not hold 
in general for other stars and elements, and it is unclear whether the same
result is valid for classical, theoretical, 1D models.

Observational scrutiny of surface convection has concentrated on the
measurement of the asymmetry of lines, which is generally quantified as
a line bisector (Kulander \& Jefferies 1966). Line bisectors have been  
analyzed in the Sun (e.g. Dravins, Lindegren \& Nordlund 1981), and  
stars across the Hertzsprung-Russell diagram (e.g. Gray 1982;  Gray \& Toner 1986;
 Gray \& Nagel 1989). The  measurement of line shifts has not received 
 the same attention (e.g. Allende Prieto \& Garc\'{\i}a L\'opez 1998; 
 Hamilton \& Lester 1999).
  We refer the reader to  Dravins (1999) for a recent review on  line
  asymmetries and shifts generated by surface convection.

Our goal is to test a new 3D model of Procyon using a large collection of
lines observed at high-resolution and high signal-to-noise ratio.
Tests employ the line shifts, asymmetries, and widths, or in other words, 
the detailed line profiles. This the first such comprehensive test of
3D hydrodynamical models for a star other than the Sun. All the calculations
described in this paper assume LTE.

\section{Observations}

The optical spectrum of Procyon  was observed from McDonald Observatory
on  January 30 and 31 (UT), 1999. We made use of the Harlan J. 
Smith 2.7m telescope and the {\it 2dcoud\'e} spectrograph
 (Tull et al. 1995), at the  focal station F1, 
delivering a FWHM resolving
 power of $R \equiv \lambda/\delta\lambda \simeq 2 \times 10^5$. Five different
 spectral setups, selected by tilting the E2 grating, 
 provided almost complete spectral coverage from 4559
to 5780 \AA, and  different exposures (up to ten for a single setup)
  were coadded to reach signal-to-noise ratios per pixel in 
the range $550-2000$.

A large number of {\it flatfield} 
exposures was acquired in order 
to correct the  pixel-to-pixel sensitivity
pattern (typically smaller than 0.5 \%) 
without degrading 
the potentially high signal-to-noise ratio of the spectra.
The individual exposures of Procyon were shorter than, or equal to, five 
minutes to keep the signal within the linear range  of
the detector, and to minimize the blurring due to variations in the Earth's
motion and instrumental drifts. 
Standard procedures (within
IRAF\footnote{IRAF is distributed  by the National Optical Astronomy
Observatories, which are  operated by the Association of Universities
for Research in  Astronomy, Inc., under cooperative agreement with the
National Science Foundation.}) were used to remove the bias level
and  scattered light,  to correct the
pixel-to-pixel variations of sensitivity, and to extract the spectra.
The spectral coverage is complete, except for 
  ten gaps spanning from 0.2 to
12.5 \AA\ wide.  For each spectral setup, about  a hundred  
Th-Ar emission lines were  used to define the (dry standard air) wavelength
scale, by fitting a 2D polynomial with an r.m.s.  at least ten times
smaller than the dispersion.
The flux distribution along  any given order was normalized by  
a polynomial.
While the normalization procedure is expected to be adequate in spectral
regions where the density of absorption lines is low, and the  lines present
are not very strong, the  limited coverage of the individual 
spectral orders ($\sim 20 $ \AA) precludes an accurate continuum tracing
in the proximity of strong lines and  in crowded spectral regions.

\begin{figure*}
\begin{center}
\includegraphics[width=10cm,angle=90]{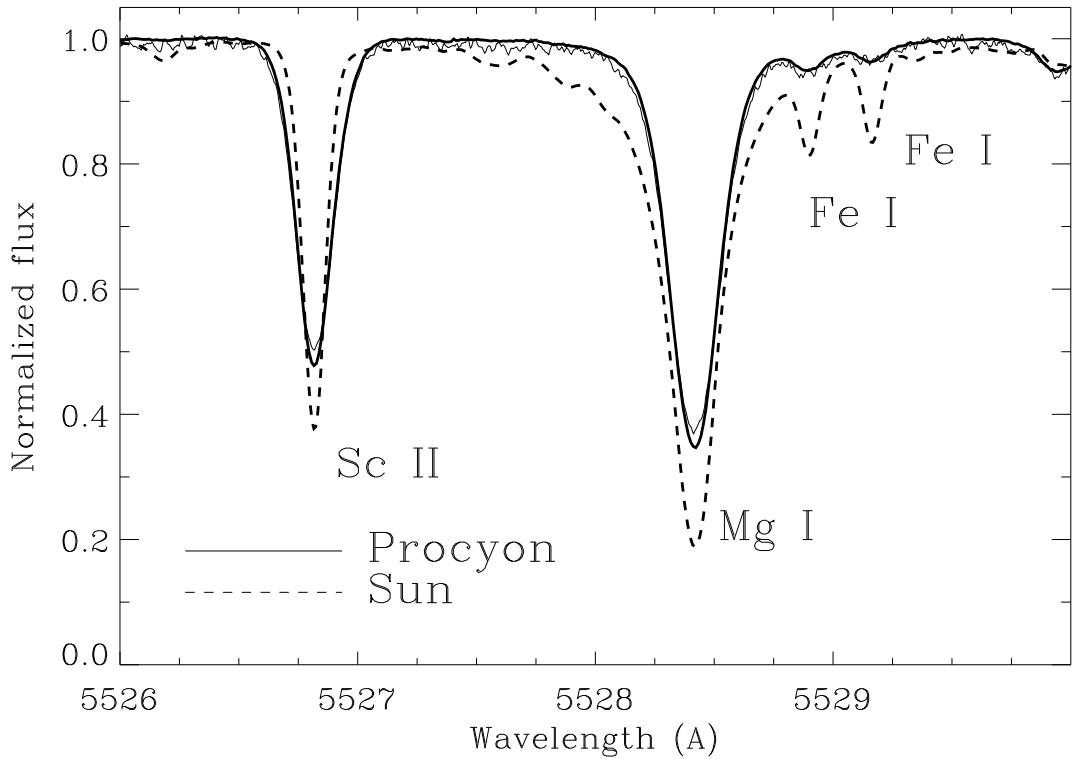}
\figcaption{
Comparison of the McDonald spectrum of Procyon (thick solid line) with the 
photographic atlas of Griffin \& Griffin (1979; thin solid line), and
the solar atlas of Kurucz et al. (1984; thick dashed line).
\label{ex}}
\end{center}
\end{figure*}

The multiorder spectra  permit a precise
correction of the velocity variations and instrumental shifts between 
different exposures, cross-correlating them order-by-order. In all cases the
standard deviation of the shifts measured between two frames from  the 16
different orders was less than 50 m s$^{-1}$, with  a mean value of 
18 m s$^{-1}$ (SEM\footnote{Standard error
of the mean: $\sigma/\sqrt{N}$}: 4.5 m s$^{-1}$). For a given spectral
setup, we shifted the individual extracted spectra with a single velocity 
to overlap them on top of the first exposure.
Then, the five different 
spectral setups were reduced to a common scale by computing and correcting the 
heliocentric velocity shifts produced by Earth's rotation,
and the motion of the Earth-Moon barycenter around the Sun. 
The extracted orders were then  combined 
 to compose a spectroscopic atlas, which spans the aforementioned
 spectral range 
 at a variable dispersion  of $0.0083 \le \Delta\lambda \le 0.0109$ \AA/pix. 
 Figure \ref{ex} displays the spectrum of Procyon in the vicinity of the
Mg I line at 5528.4 \AA, and compares it with the solar spectrum (Kurucz,
Furenlid \& Brault 1984) and the photographic atlas of Procyon by Griffin \& 
Griffin (1979). The new data are publicly available through the 
internet\footnote{{\tt http://hebe.as.utexas.edu/procyon/}}.

\begin{figure*}
\begin{center}
\includegraphics[width=14cm,angle=0]{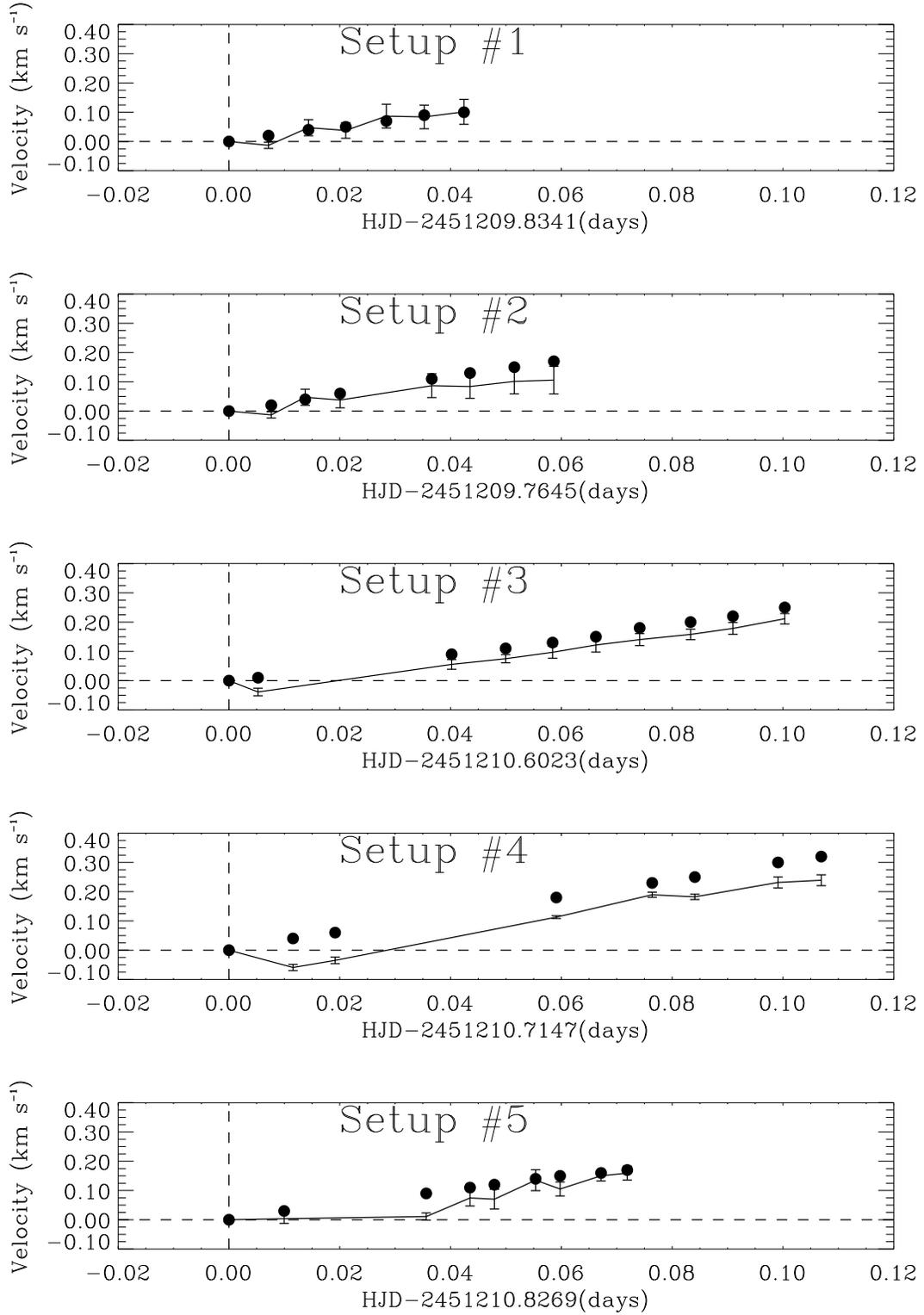}
\figcaption{
Velocity shifts determined from the cross-correlation of the individual
exposures for each setup (solid line with error bars). The error bars are
determined as the standard error of the mean from the scatter for all
the spectral orders. The velocity shifts produced by Earth's motion 
are marked with filled circles. All shifts are relative to the first spectrum 
in each series.
\label{drift}}
\end{center}
\end{figure*}

\begin{figure*}
\begin{center}
\includegraphics[width=7cm,angle=90]{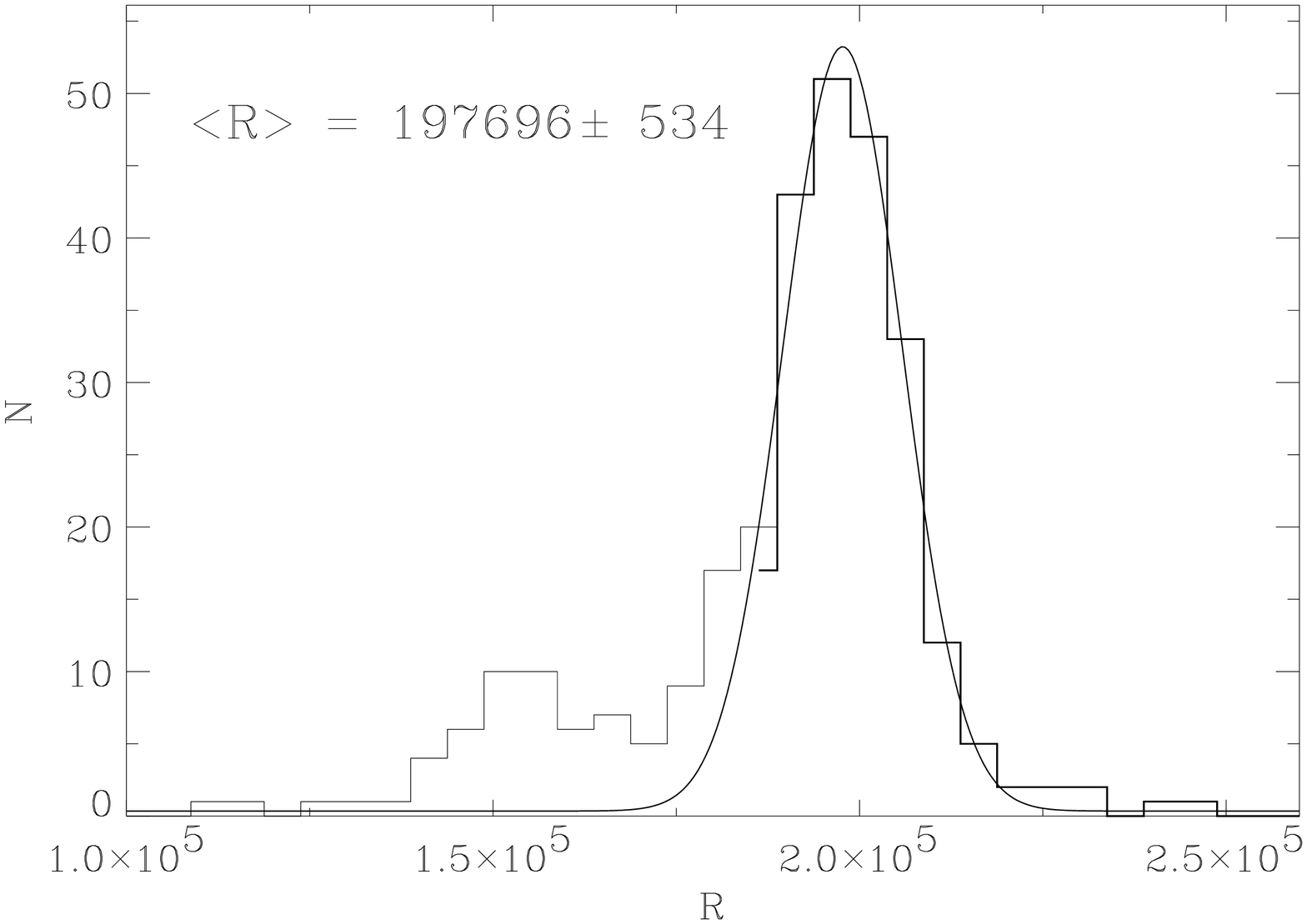}
\figcaption{
Number of emission lines measured in the Th-Ar hollow cathode spectra
as a function of the resolving power $R \equiv \lambda/\delta\lambda$. Cosmic
rays are easily identified and excluded from the analysis. 
 Systematic
 errors from blended features are minimized by fitting only the right
 flank of the nearly Gaussian distribution (thick part of the histogram).
\label{resolution}}
\end{center}
\end{figure*}

In a work of this nature, it is very important to check for possible 
instrumental
effects that might distort the  stellar line profiles. The 
{\it 2dcoud\'e} spectrograph is a well-tested
instrument but instrumental drifts have been suspected (see, e.g., Allende 
Prieto et al. 1999a).
  Figure \ref{drift} shows the variation of the heliocentric correction 
  for the 
 individual exposures respect to the first (filled circles), 
and the velocity
 drifts determined from cross-correlating the individual exposures with the
 first as template. 
A few relatively large 
  drifts (up to 50 m s$^{-1}$) are present between consecutive exposures, 
  but slower drifts are also
 obvious in some cases.  Spectra of the
 Th-Ar hollow cathode were acquired immediately before and after each stellar
 exposure. The reference wavelengths for the Th-Ar lines used in the 
 calibration 
 correspond to dry standard air (760 mmHg and 15 $^{\rm o}$C), and the
 profusion of calibration spectra should 
 prevent large drifts as pressure, temperature, and humidity drag the 
 air refraction index.  Nonetheless, the sensitivity to changes in temperature, 
 pressure
 and humidity is such that variations on small time scales -- between 
 the  exposures of the Th-Ar hollow cathode and the star --  can definitely 
 be a source of small errors. 
  For example, a change of 0.1 $^{\rm o}$C from standard conditions would induce a change 
 of 28 m s$^{-1}$, a change of 0.1 \% in pressure would induce a 
 drift of 82 m s$^{-1}$, and introducing 0.1 \% of  water vapor
 would produce a change of 13 m s$^{-1}$ at 5000 \AA. Lacking a proper study of
 these effects in our spectra, we note that changes in pressure 
 within the spectrograph
 follow in full amplitude those outside the building, and variations as large as
 than 0.07 \% per hour were measured on the observing dates. Additionally,
it is known that the position of the CCD varies with the weight of
the N$_2$ dewar.
Differences between how 
 the slit is illuminated by the Th-Ar hollow cathode and the stellar light cannot
be  ruled out as contributor to systematic errors. Although the 
cross-correlation technique makes it possible to determine
the shifts among the individual exposures for a given setup to within 
$\simeq 5$  m s$^{-1}$, the observed systematic effects introduce an additional 
uncertainty in our wavelength calibration, probably not larger than 20--30 
m s$^{-1}$. Considering more than 200 lines in the Th-Ar spectra, 
we confirmed that 
the nominal resolving power was indeed 
achieved ($R = 197700 \pm 500$; see Fig. \ref{resolution}), with no detectable
variation with order or wavelength, but a small dependence on the  
setup.

\section{Observed signatures of convection}

Different factors may contribute to the observed line asymmetries and 
shifts. Blends with other features affect particular lines in different 
ways that can not be generally predicted with precision.
 Therefore, it is necessary to average  observed line
asymmetries and shifts for a large number of lines in order to 
quantify the signatures of convection with minimal distortion from
other  shifts. Convective 
line asymmetries and shifts are closely related and  carry
complementary information. However, the limited spectral range 
 of earlier studies 
 prevented the  study of  line shifts in depth.
In short,  stellar studies have ignored line shifts and relied on
 mean line bisectors computed assuming that
the velocity of the lowest end of the bisector is the same for all lines.  

Previous studies of line asymmetries in Procyon's spectrum were limited
in scope.  Gray (1981a, 1982) analyzed lines in a  very small spectral
window ($\sim 70$ \AA) acquired at  $R=1.2 \times 10^{5}$.  
Rice \& Wehlau (1984) attempted 
the observation but at  insufficient spectral resolution.
Dravins (1987) analyzed 11 lines using  
the  ESO coud\'e spectrometer double-pass photoelectric scanner at 
$R=2 \times 10^{5}$, and
 he strengthened his results by demonstrating that, despite  limited
signal-to-noise ratio, the scanned version of the photographic atlas of
Griffin \& Griffin (1979) could be used  to
derive  statistically-meaningful averaged bisectors for sets of  lines with
different strengths.  Our much larger spectral coverage than 
 previous CCD studies,  higher
signal-to-noise ratio and resolution,
 and thorough control of possible systematic effects,
  set the stage for  the first
detailed study of asymmetries and shifts 
 for a large sample of iron lines.  The
gravitational redshift and the orbital and radial velocities affect
 the zero point of the velocity scale, but the information we are searching for
  is contained in the line bisectors and their relative displacements.

To minimize  systematic differences in the comparison with the Sun, we
have carried out  an identical analysis with the solar  atlas of Kurucz 
et al. (1984). The lines were selected (or rejected) and
classified following the same criteria for both stars, in an attempt to
keep the comparison as meaningful as possible. It is of great interest
to analyze the  asymmetries of Fe I lines in the flux spectrum of the
Sun on an absolute scale, as this has never been done consistently
before, and must provide the best standard for  comparisons with other
stars. Finally, even at risk of repeating what has already been  said 
many times in the literature, we note that there are several good reasons 
to pick Fe I among the observed species: the large number of  
lines in late-type stars, the lack of measurable isotopic shifts and hyperfine
splitting, and the availability of accurate laboratory wavelengths.

\subsection{Line shifts}
\label{ls}

We used the line list of Th\'evenin (1989, 1990)  to search for
Fe I lines in the solar spectrum. The central wavelengths 
and bisectors of the Fe I lines in
that list  were measured in the solar atlas  
and the McDonald spectrum of Procyon, and then compared with the 
precise laboratory
wavelengths measured by Nave et al. (1994). The 
wavelength calibration of the solar atlas has been examined by 
Allende Prieto \& Garc\'{\i}a L\'opez (1998), who showed 
that it is highly reliable.

The measurement of the central wavelength of the line profiles was
carried out fitting a third-order polynomial to the 11 lowest points
around the minimum, equivalent to an interval of $90-120$ m\AA\ in the
spectrum of Procyon and 50 m\AA\ in the solar spectrum. 
 We measured all the  lines in Th\'evenin's list  identified
as Fe I lines.
 For this comparison, the equivalent widths of the lines were estimated 
 by synthesizing the
lines with Kurucz (1992) model  atmospheres corresponding to the Sun and Procyon. This is safer 
than attempting to measure the equivalent
widths in the spectra, as a large fraction of the lines are strongly
blended outside their core, especially at the bluer wavelengths. We
made use of the {\it solar} $gf-$values of Th\'evenin (1989, 1990) and 
the damping enhancement factors for Fe I recommended by the
Blackwell group, as implemented in the LTE synthesis program MOOG
(Sneden 1973).

\begin{figure*}
\begin{center}
\includegraphics[width=10cm,angle=90]{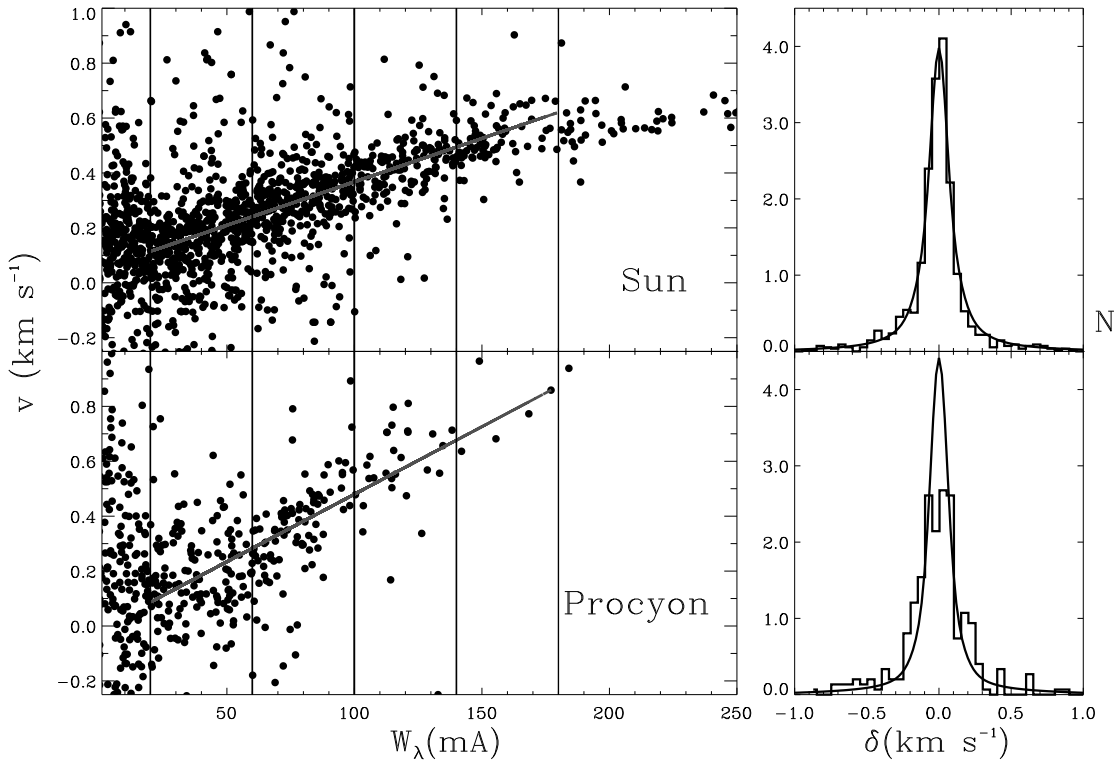}
\figcaption{
Left panels: velocity shifts between the stellar and laboratory 
wavelengths for relatively weak lines; 
for Procyon the zero-point
arbitrary. The strong solid line is a least-squares linear fit for 
all lines with $20 \le W_{\lambda}
\le 180$ m\AA. 
Right panels: difference between the
measured shifts and the linear fits to lines with $20 \le W_{\lambda}
\le 180$ m\AA\ (histogram); the expected distribution,
based on error estimates for the laboratory wavelengths and the
polynomial fits to the center of the stellar lines, is also shown (solid 
 curve).
\label{shifts_low}}
\end{center}
\end{figure*}

The left panels in Fig. \ref{shifts_low} show the Fe I line velocity 
shifts (v) measured in the spectra of 
 the Sun and Procyon.   The smaller  wavelength coverage 
available for Procyon (4559 to 5780 \AA)  in comparison with the solar
atlas of Kurucz et al. (2960 to 13,000 \AA), reduces the number of 
measured lines from 1551 to 506. The line shifts exhibit similar 
patterns for both stars. There is no reason to think that the 
well-known  granulation observed in the Sun is not 
present in  Procyon, and we naturally associate the larger
 shifts towards the red for stronger lines with a decreasing  of the 
convective blue-shifts. Unfortunately, there is no line in  the observed part 
of Procyon's spectrum stronger than 200 m\AA, 
leaving open the question of whether the 
velocity shifts of the strongest Fe I lines fall on a solar-like 'plateau'. 
It is interesting to compare the left-hand panels 
in Fig. \ref{shifts_low}. The
velocity shifts in the solar spectrum span a range of 
$\sim 0.6$ km s$^{-1}$,
but at least 0.9 km s$^{-1}$ for Procyon's spectrum.

The velocity indicated by the strong lines on the solar 
plateau ($W_{\lambda}  \ge 200$ m\AA) is essentially 
independent of the line's equivalent width.
We note that Fig. \ref{shifts_low} 
has been truncated at $W_{\lambda} = 250 $ m\AA, but the solar plateau
extends at least to lines  as strong as 2 \AA\ (see
 Allende Prieto \& Garc\'{\i}a L\'opez 1998). Part of the observed scatter 
 should be 
attributable to the laboratory wavelengths and uncertainties 
in the measurement 
of the center of a spectral line (with a finite width), which 
can be quantified. Nave et al. (1994) classify their wavelengths into four 
categories,
depending on their uncertainty ($\sigma_{\rm l}$), which ranges 
from 0.4 m\AA\ to more than 
10 m\AA. Most lines have errors below 5 m\AA\ and about 
half of the 
lines have errors below 1 m\AA. Errors in the measured 
stellar wavelengths 
can be estimated  from the horizontal scatter of the 
polynomial least-squares fit to 
the line center divided by the square root of the number of points in the
fit (in this case 11), $\sigma_{\rm s}$. 

Neglecting the effect of unidentified line blends and systematic effects 
in the wavelength scales of the stellar spectra, the 
uncertainties in the measured  line shifts, 
$\sigma = \sqrt{\sigma_{\rm s}^2 + \sigma_{\rm l}^2}$, or more precisely, 
the  distribution of
uncertainties, $n(\sigma)$, can  be used to predict the distribution of 
errors ($\delta$) in the observed velocity shifts  as

\begin{equation}
N(\delta) = \frac{1}{C} \int_{0}^{\infty}  \frac{n(\sigma)}{\sigma} 
\exp \left(-\frac{\delta^2}{2\sigma^2}\right) d\sigma,
\label{ndelta}
\end{equation}

\noindent where $C$ is chosen such  that 
$\int_{-\infty}^{\infty} N(\delta) d\delta = 1$. The derived $N(\delta)$ can be 
compared to the observations 
in order to constrain the intrinsic scatter.

Discarding a deviant line, the average shift for the 19 measured Fe I solar lines 
stronger  than 400 m\AA\ is 
${\rm v} = 0.626 \pm 0.011$ (SEM) km s$^{-1}$, consistent with the gravitational 
redshift of solar photospheric light at Earth, 0.633 km s$^{-1}$. 
The strengthening of the convective blueshifts for weaker lines 
is definitely not far from linear. A 
linear regression for lines with  equivalent widths $20 \le W_{\lambda} 
\le 180$ m\AA\ provides, in the solar case,  a similar slope as linear 
fits within several  subintervals included in that range. Conversely, 
there is some marginal indication that the trend flattens for the lines 
with $20 \le W_{\lambda} \le 60$ m\AA\ in the spectrum of Procyon. 
The intrinsic scatter for the Sun is very small; at least significantly smaller
than our measurement errors (see Fig. \ref{shifts_low}). For Procyon, 
there is an additional 
 scatter, but it should be kept in mind that we have neglected other sources
of systematic errors, such as those in the theoretical  equivalent 
widths. We will return to this issue in the light of 3D model atmospheres.

Making use of the laboratory wavelengths 
measured recently by S. Johansson (private communication), we derive the
velocity shifts for 25 Fe II lines, which exhibit a trend similar to that  
for Fe I lines. The velocity shifts of Fe II lines in the 
spectrum of the Sun are generally more shifted to the blue than 
the Fe I lines, in agreement with
the results obtained by Dravins, Larsson, \& Nordlund (1986) for
the center of the disk, but the difference becomes indistinguishable
 in the spectrum of Procyon.

\subsection{Line asymmetries}
\label{la}

We have tried to avoid subjective criteria in selecting  lines for 
measuring bisectors.  
Knowing that average convective line asymmetries 
for main-sequence F-K stars 
have  been neither measured nor predicted to be larger 
than a few hundreds  of meters per second, we have  kept only those
 lines whose  bisectors  do not exceed 1 km s$^{-1}$ in amplitude.
The very large number of available lines
in the spectrum whose rest wavelength is reliably known, makes it 
possible to use such a strong rejection condition. 
In addition, only lines whose absorption was deeper than  10 \% 
of the continuum level were included in the analysis. 
These criteria were satisfied by 331 lines in the solar atlas and by 151
in Procyon's spectra. They were
ordered in four different  groups depending on their continuum normalized 
 flux at the line center $h$:  $h < 0.35$, $0.35 <  h  < 0.55$, 
 $0.55 < h < 0.75$, 
and $h > 0.75$. Their bisectors were
 averaged retaining the relative velocity differences between
 the lowest end of the individual bisectors. In the 
process of averaging we allow for  a {\it cleaning}, in the sense that at
each  absorption depth at which the bisectors were measured, we reject
those  deviating by more than 2$\sigma$ from the mean.

\begin{figure*}
\begin{center}
\includegraphics[width=7cm,angle=90]{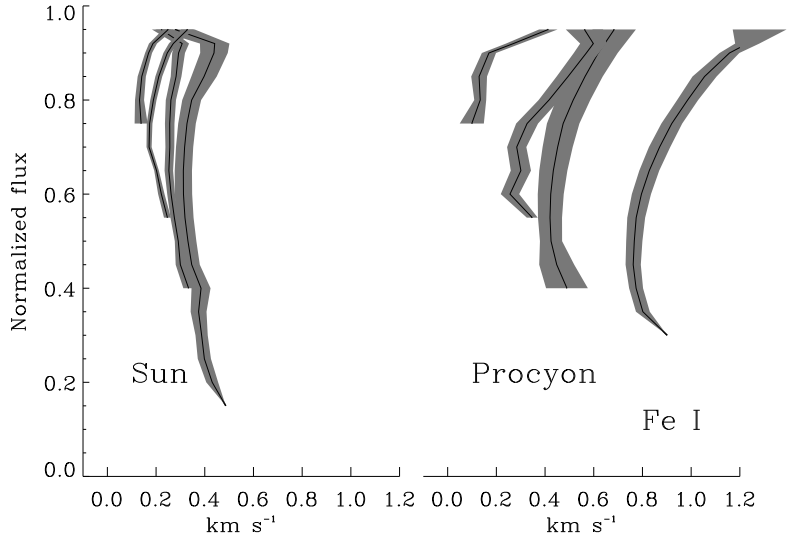}
\figcaption{
Average bisectors for Fe I lines with a continuum-normalized flux at 
the line center f: $h < 0.35$, $0.35 < h < 0.55$, $0.55 < h < 0.75$, 
and $h> 0.75$.
\label{bisectors}}
\end{center}
\end{figure*}

Fig. \ref{bisectors} shows the averaged bisectors for Fe I lines. 
The solar bisectors
are similar  to the averaged line bisectors measured in 
the spectrum of the center of the disk
 by Dravins, Lindegren \& Nordlund (1981). 
The typical 'C' shape is clearly visible. A comparison of the velocity
 spans for the Sun and Procyon
 is in  agreement with previous results:   the span
of Procyon's lines exceeds the solar values  by more than a factor of two.  
A similar result holds for the line shifts.
As  explained above, the velocity differences between the individual bisectors
were retained for averaging, but we have not subtracted the gravitational 
redshift in the solar case and the zero velocity for Procyon's bisector
was arbitrarily set.

\begin{figure*}[t!]
\begin{center}
\includegraphics[width=7cm,angle=90]{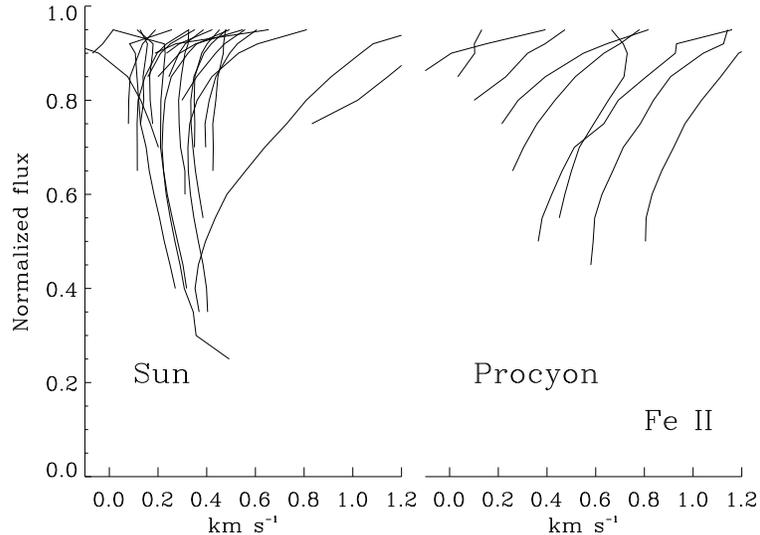}\figcaption{
Individual bisectors for Fe II lines.
\label{bisectorsII}}
\end{center}
\end{figure*}

The bisectors measured for Fe II lines are
displayed in Fig. \ref{bisectorsII}. As they are only a few, we did not average them
out. Three solar bisectors (not shown) were rejected by the 2$\sigma$ 
criterion, and there are still two more obvious outliers kept.
 Fe II bisectors corroborate the qualitative conclusions  from Fe I lines.

\section{Modeling}

\subsection{Fundamental stellar parameters}
\label{sp}

Procyon A is in a 40-year period {\it visual} binary system, sharing the
systemic velocity with a white dwarf. Girard et al. (2000) have updated
its astrometric orbit from photographic plates obtained at different
observatories between 1912 and 1995, and made a direct measurement of
the  angular separation of the pair using the infrared cold coronagraph (CoCo)
at the NASA Infrared Telescope Facility and the WFPC2 onboard HST. At a 
distance of only 3.5 pc from the Sun, the parallax measured by {\it
Hipparcos} is very precise: $p = (285.93 \pm 0.88) \times 10^{-3}$ arcsec. In addition,
Mozurkewich et al. (1991) measured the stellar angular diameter using
optical interferometry: $\theta = (5.51 \pm 0.05) \times 10^{-3} $ arcsec.  
Therefore, its
mass  can be  constrained to

\begin{equation}
M = \frac{\alpha}{p^2 P^2} \left(1 - \frac{\alpha_A}{\alpha} \right) = 1.42 \pm 0.06 M_{\odot}, 
\end{equation}

\noindent where $\alpha$, $\alpha_A$, and $P$ are the semi-major 
axis of the visual orbit or Procyon A, the semi-major axis of its orbit 
relative to the  barycenter of the system, and the orbital period, respectively, and
for which we have adopted $\alpha= 4.271 \pm 0.032 $ arcsec, 
$\alpha_A = 1.232 \pm 0.08$ arcsec, and $P=40.82 \pm 0.06$ yr. 
 (Girard et al. 2000),  in combination with {\it Hipparcos}'s parallax.  
The radius is

\begin{equation}
R \simeq \frac{\theta}{2 p \tan (\frac{\theta_{\odot}}{2})} = 
2.071 \pm 0.020 R_{\odot}, 
\end{equation}

\noindent where we have adopted 
$\theta_{\odot}/2 = 959.64 \pm 0.02 $ arcsec (Chollet \& Sinceac 1999), 
leading to a gravity $\log g = 3.96 \pm 0.02$ (c.g.s. units). 
The derived gravity is in good agreement with the estimate of Allende Prieto
\& Lambert (1999), $\log g = 4.04 \pm 0.14$ dex, obtained from 
matching the observed $M_V$ and $(B-V)$ to the evolutionary models of
Bertelli et al. (1994)\footnote{We note, however, that the estimate 
in a previous paper by Allende Prieto et al. (1999b) for this and other stars
in their sample with $T_{\rm eff} > 6000$ K  is underestimated, as a result of
  the improper use of a  different set of isochrones restricted to ages 
  larger than 5 Gyr.}.
Our mass estimate is slightly lower from that 
given in Girard et al. (2000), as we prefer to adopt the {\it Hipparcos} 
parallax and 
the angular separation determined from  HST-WFPC2 observations, 
discarding  the   ground-based measurement. 
Adopting the parallax derived 
by Girard et al., $p = (283.2 \pm 1.5) \times 10^{-3}$ arcsec, 
we find only a minor correction 
to the derived mass: $M= 1.47 \pm 0.06 M_{\odot}$.

The star has a Johnson visual magnitude of $V = 0.363 \pm 0.003$ mag
(and therefore  an absolute magnitude $M_V = 2.644 \pm 0.007$) and a
color index $B-V = 0.421 \pm 0.003$ mag, as derived from 13
measurements extracted from SIMBAD. Fuhrmann et al. (1997) compiled
several estimates of Procyon's bolometric flux from the literature.  Their 
weighted average is $F_{\rm BOL} = (18.20 \pm 0.43) \times 10^{-6}$ 
erg cm$^{-2}$ s$^{-1}$ and, therefore,
 the star has an effective temperature 

\begin{equation}
T_{\rm eff} = 7.400 \times 10^3
\left(\frac{F_{\rm BOL}}{\theta^2}\right)^{1/4} = 6512 \pm 49 ~ {\rm K}.  
\end{equation}

\noindent Finally, a
review of the high-resolution spectroscopic analyses previously
performed, indicates that the star has either  solar,  or slightly
lower than solar,  iron abundance.

Linear interpolation in the solar-metallicity isochrones published by
Bertelli et al. (1994) to match the pair ($M_V$,$B-V$)  constrains very
tightly the age of Procyon A to the range $1.66-1.73$ Gyr,
and predicts a gravity in agreement with the astrometric value. 
Alternatively, matching the observed radius and mass would  shift the
age estimate to $2.0$ Gyr. 
This last value, less affected by
uncertainties in the conversion between the theoretical and
observational quantities, is to be preferred.  These ages are significantly 
shorter than the nuclear time-scale of 3.9 Gyr  adopted by 
Provencal et al. (1997),  
implying a  mass for the progenitor of the white dwarf (Procyon B)
that is larger than previously estimated. 

\subsection{Model atmospheres}
\label{ma}

Realistic {\it ab-initio}, compressible, 
radiative-hydrodynamical simulations of  
surface convection in Procyon have been performed with the  code 
   formerly used to model solar 
(e.g. Stein \& Nordlund 1998; Asplund et al. 2000a,b,c; Asplund 2000)
and stellar granulation (Asplund et al. 1999; 
Asplund \& Garc\'{\i}a P{\'e}rez 2001).
The resulting structure is a 3D, time-dependent
model atmosphere with a self-consistent description of the convective
flows. The hydrodynamical equations of mass, 
momentum and energy conservation:

\begin{equation}
\frac{\partial {\rm ln} \rho}{\partial t} = - {\bf \bar{v}} \cdot \nabla 
{\rm ln} \rho - \nabla \cdot {\bf \bar{v}}
\end{equation}

\begin{equation}
\frac{\partial {\bf \bar{v}}}{\partial t} = - {\bf \bar{v}} \cdot \nabla 
{\bf \bar{v}} + {\bf \bar{g}}
- \frac{P}{\rho} \nabla {\rm ln} P + \frac{1}{\rho} \nabla \cdot \sigma
\end{equation}

\begin{equation}
\frac{\partial e}{\partial t} = - {\bf \bar{v}} \cdot \nabla e 
- \frac{P}{\rho} \nabla \cdot {\bf \bar{v}} + Q_{\rm rad} + Q_{\rm visc}
\end{equation}

\noindent coupled to the equation of radiative transfer:

\begin{equation}
 \left({\bf n} \cdot {\bf \nabla}\right)  I_{\lambda} = \eta_{\lambda} - \kappa_{\lambda} I_{\lambda}
\end{equation}

\noindent have been solved on a non-staggered Eulerian mesh with 
$100 \times 100 \times  82$ grid-points.
In the above equations, $\rho$ denotes the density,
${\bf \bar{v}}$ the velocity, ${\bf \bar{g}}$ the gravitational acceleration,
$P$ the pressure, $e$ the internal energy,
$\sigma$ the viscous stress tensor, $Q_{\rm visc}$ the viscous dissipation,
$Q_{\rm rad} = \int_\lambda \int_\Omega \kappa_\lambda 
(I_\lambda-S_\lambda){\rm d}\Omega{\rm d}\lambda$ 
the radiative heating/cooling rate, $I_\lambda$ the monochromatic intensity,
with  $\kappa_{\lambda}$ and $\eta \equiv B_\lambda(T) \kappa_{\lambda}$ 
representing the absorption and emission 
coefficients, respectively, where $B_\lambda(T)$ is the Planck function,
and ${\bf n}$ the unit vector for each of the eight
considered  directions. 
The code was stabilized
using a hyper-viscosity diffusion algorithm (Stein \& Nordlund 1998) with
the parameters determined from standard hydrodynamical test cases, like
the shock tube.
Periodic horizontal boundary conditions and open transmitting top
and bottom boundaries were used in an identical fashion to
the simulations described by Stein \& Nordlund.

In order to make the atmospheric structure as realistic as possible for 
direct confrontation with observations, we used a detailed 
 equation-of-state (Mihalas et al. 1988), including 
 the effects of ionization, excitation and dissociation, as well as 
   continuum  (Uppsala opacity package; see Gustafsson et al. 1975;
Asplund et al. 1997) and line (Kurucz 1993) opacities. 
The 3D radiative transfer has been solved 
at each time-step of the simulation by Feautrier's method, under the 
approximations of 
local thermodynamic equilibrium (LTE) and the opacity binning technique 
(Nordlund 1982). The accuracy of the opacity binning procedure has been
verified at regular intervals by solving the full monochromatic radiative
transfer (2748 wavelength points) in the 1.5D approximation, i.e., treating 
each vertical column separately and thus ignoring  horizontal radiative
transfer effects.  

The physical dimension of the numerical box is 
$21 \times 21\times 13$ Mm, of which about 3\,Mm correspond to the
 photosphere. A
snapshot from a previous lower-resolution simulation sequence of 
Procyon with a less extended depth-scale (Trampedach 1997) was used 
as  initial condition. The full convection simulation covers more
than 3 hours of stellar time, but only the final hour has been used
for the calculations of  spectral line formation presented here. 
The resulting effective temperature  of the 1 hour sequence
was $6514\pm27$\,K, in close agreement with the interferometric estimate. 
We adopted a surface gravity of log\,$g=3.96$\, (cgs)  and 
solar chemical composition (Grevesse \& Sauval 1998).
In particular, the Fe abundance
used for the equation-of-state and continuum opacity calculations was 
$\log \epsilon_{\rm Fe}= 7.50$\footnote{On the customary logarithmic 
abundance scale defined
to have log$\,\epsilon_{\rm H}=12.00$}. The line opacities were 
extrapolated to the adopted Fe abundance using the standard Kurucz ODFs with  
log$\,\epsilon_{\rm Fe}=7.67$ and non-standard ODFs computed with 
log$\,\epsilon_{\rm Fe}=7.51$ and with no He
(Kurucz 1997, private communication, cf. Trampedach 1997 for details
of the procedure). 

\begin{figure*}
\begin{center}
\includegraphics[width=20cm,angle=0]{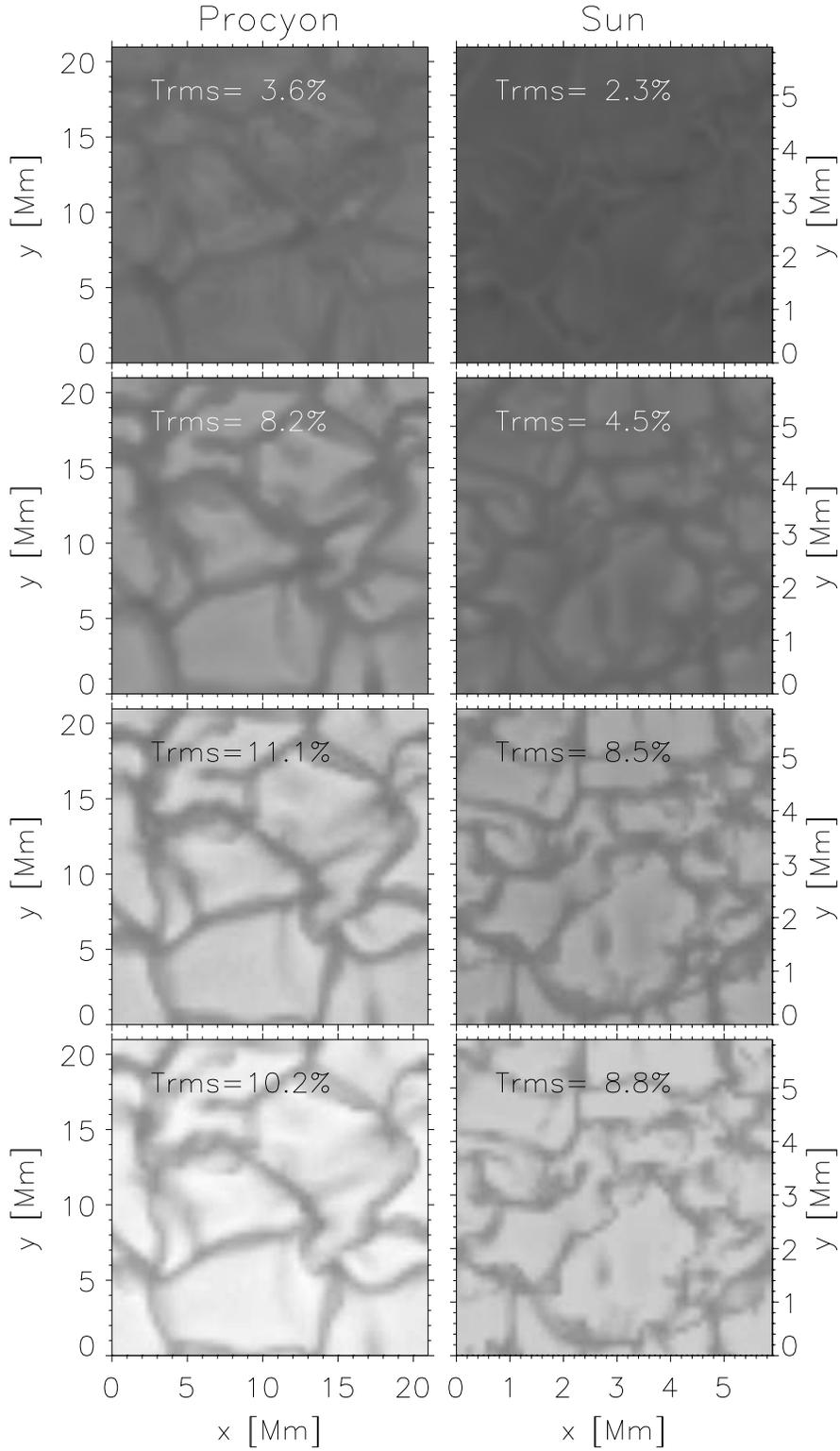}
\figcaption{
The temperature on surfaces of equal optical depths in the convection
simulations of Procyon (left panel) and the Sun (right panel). 
The surfaces correspond to continuum optical depths 
at 5000 \AA\ of 0.3, 1, 3 and 10 from top to bottom panels. It should be noted
that these iso-tau surfaces are highly corrugated and therefore the temperature
contrast is much greater across surfaces of equal geometrical depths. 
All images share the same maximum and minimum temperature cuts, 
which highlights the significantly 
larger temperature contrast in Procyon. For each iso-tau surface, the 
temperature contrast ($T_{\rm rms}/\langle T \rangle$) is given.
\label{ttau}}
\end{center}
\end{figure*}

The resulting convective structures for Procyon
are qualitatively similar to the lower resolution runs 
presented in Nordlund \& Dravins (1990),
which, however, were computed with the anelastic approximation and 
less complete equation of state and opacities.
With the new compressible code, larger convective velocities 
are encountered, and they often exceed the speed of sound, causing prominent
shocks in the photosphere.
The phenomenon of "naked granulation" discussed by Nordlund \& Dravins (1990)
is also present in the improved simulations as seen in Fig. \ref{ttau}:
the layers with largest
temperature contrast due to the convective motions are located around the
continuum forming region ($\tau_{5000} \simeq 1-3$)
and not slightly beneath it, as is the case with
cooler stars like the Sun ($\tau_{5000} \simeq 10$). 
In addition, the temperature contrast itself is significantly larger in
Procyon than in the Sun.
All these features  manifest themselves
in more pronounced line asymmetries. 
A detailed description of the convective structures
in the granulation simulations of Procyon and other solar-type stars
will be presented in a subsequent article,
while here we concentrate on the effects upon spectral line formation
and the confrontation with observations.

To enable a strictly differential comparison between the 3D and 1D predictions
of the resulting line formation, 1D, hydrostatic, plane-parallel model atmospheres
have been generated with the {\sc marcs} code (Asplund et al. 1997). 
These homogeneous models are fully line-blanketed
and have identical defining parameters as the 3D simulations, including 
all chemical abundances. 

\subsection{Spectral line calculations}

Using the granulation snapshots of the Procyon simulation as
3D model atmospheres, the spectral line formation was computed
for a sample of Fe I and Fe II lines  assuming
 LTE ($S_\nu = B_\nu$). The original
simulation data was interpolated to a denser vertical grid
covering only the outermost 6\,Mm  but 
with same number of
depth points as before. 
From the stored temperatures and densities of the snapshots,
monochromatic continuous opacities were computed using the 
Uppsala opacity package,
which was also 
employed for constructing the 1D model atmospheres we compare with. 
The assumptions of Boltzmann and Saha distributions and chemical 
equilibrium have been made in 
the calculation of opacities and number densities of the various
species.

The radiative transfer was solved for five center-to-limb positions
and seven equally spaced azimuthal angles, for a total of
29 different directions. The Gaussian distributed $\mu$-angles were
chosen to facilitate the disk integration 
taking into account the rotational velocity of the star
(Dravins \& Nordlund 1990). 

The flux profiles at different velocities were computed as:

\begin{equation}
F(v) = \int^{2\pi}_{\varphi = 0} \int^{1}_{\mu = 0} 
I(v-v_{\rm rot}\sin i \sin\theta \cos\varphi, \mu,\varphi)
\mu{\rm d}\mu{\rm d}\varphi,
\end{equation}

\noindent with $\mu= \cos \theta$. The integrals were evaluated using a four-point Gaussian 
quadrature for $\mu$ and seven equidistant $\varphi$ angles with
equal weight. Finally, the rotationally broadened profiles for the
31 individual snapshots were added together before normalization
to the local continuum. It was verified through
test calculations that using twice as many $\mu$- and $\varphi$-angles
results in insignificant differences in the theoretical line asymmetries.
The procedure  accounts for the Doppler shifts introduced
by the convective motions in the stellar atmosphere which, together with 
the temperature inhomogeneities, produce the characteristic line bisector shapes.
Various tests showed that the temporal coverage of the convection
simulation produces statistically meaningful 
line profiles and asymmetries. 
Following the calculations of disk-integration and rotational broadening,
the resulting profiles were convolved with a Gaussian with a
FWHM of 1.5\,km\,s$^{-1}$,
to account for the finite spectral resolution of the spectrograph. 
As shown for the Sun, for which the rotational velocity is accurately known,
there is no need to introduce any additional line broadening in the form
of micro- and macroturbulence in analyses based on high-resolution
3D model atmospheres 
(Asplund et al. 2000a,b,c; Asplund 2000).

The profiles were computed for 141 equidistant velocities, covering
$-28$ to $+28$\,km\,s$^{-1}$ around the laboratory wavelengths.  Each
line was calculated for three different Fe abundances, 7.20, 7.50 and
7.80, from which the final profile was interpolated. 
Test calculations verified that the abundance
interval was sufficiently small not to introduce any systematic errors
in the abundances ($<0.01$\,dex) and line asymmetries ($<0.02$\,km\,s$^{-1}$) 
derived.

The 1D spectral line calculations for the comparison with the 3D
results have been computed with the same spectral synthesis code as the
3D profiles. 
Without the convective Doppler shifts, additional adhoc
broadening in the form of the micro- and macroturbulence must
be invoked in 1D in order to obtain correct line widths, with the former
affecting the line strengths and the latter only the line shapes. 
In both cases, Gaussian distributions are assumed.

\subsection{Line data}
\label{gfs}

Iron is the best represented element in the spectrum of late-type
stars. Neutral iron has been the subject of a number of fine
laboratory works to derive radiative transition probabilities at Oxford 
(e.g. Blackwell et al. 1986). The number of lines measured with
high accuracy and in a homogeneous manner has  been  
enlarged by the work of O'Brian et al. (1991).  Shown to be in 
good agreement with the Oxford scale (see Lambert et al. 1996), 
we have adopted this last reference. Unfortunately,
the situation for ionized iron is worse. To select transition probabilities 
for this species, we have  restricted the search to laboratory
measurements. Theoretical calculations  still show many systematic
effects as well as a large scatter. Astrophysical determinations always
imply a certain prejudice that we wish to avoid: a selection of a model
atmosphere, abundances, collisional damping parameters, etc.  
As we felt that an updated
critical compilation of laboratory  Fe II $f$-values was lacking, we have
looked for data for all lines with wavelengths longer than 300 nm, 
rather than
restricting the search to the spectral range of  Procyon's
observations. The method  adopted to average the Fe II
transition probabilities and the data are described in the
Appendix. By combining laboratory, {\it astrophysical}, 
and theoretical  determinations, it is  possible to amass a larger 
line list, but at the expense of 
higher uncertainties (see, e.g., Giridhar \& Arellano Ferro 1995).

Whenever possible,  new quantum-mechanical calculations
(Anstee \& O'Mara 1991; Barklem \& O'Mara 1997; Barklem et al. 1998, 2000)
have been employed for the pressure broadening of the spectral lines
by collisions with neutral hydrogen.
Otherwise (in particular for the Fe II  lines as the quantum-mechanical 
treatment has not yet been generalized to ionized species),
the classical formula proposed by Uns\"old (1955), 
with an additional enhancement factor $E=2.0$, 
has been adopted. Radiative damping was included with 
values obtained from VALD (Kupka et al. 1999).
Stark broadening was not considered.

We selected lines of neutral and singly-ionized iron  from the
observed spectrum that were included in the list of O'Brian
et al. (Fe I), or in our list in the Appendix (Fe II).  Many lines
were eliminated from a preliminary list, since they were apparently 
blended with
other features. Visual inspection decided which lines and parts of the profiles
were selected at this point. The lines are listed in Table 1.

In addition, we extracted the profiles of several Fe I and Fe II lines
 from the solar atlas of Kurucz et al. (1984), in order to 
estimate the solar iron abundance in a similar fashion.
The lines 
  were selected from the list in Asplund et al. (2000c).
  They show clean profiles, and the sources for the atomic data 
    are identical to those used for Procyon. These lines 
    and their atomic data appear in Table 2. In some cases,
 the transition probabilities  differ from those given in 
Asplund et al. (2000c); the difference
is however marginal ($\simeq 0.02$ dex on average for both Fe I and Fe II).
We noticed some solar Fe lines partially
 affected by blends, and the afflicted
parts of the profiles were ignored 
in the analysis.

\section{Comparison between theoretical and observed line profiles}

\subsection{Line profiles and stellar rotation}
\label{lpsr}

\begin{figure*}
\begin{center}
\includegraphics[width=12cm,angle=0]{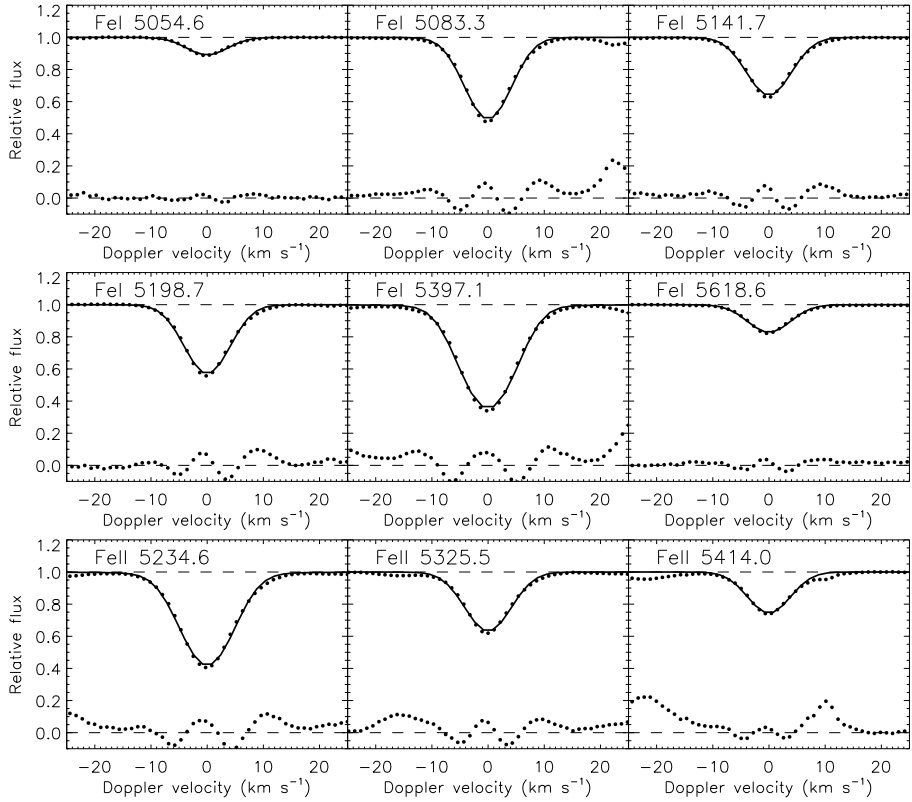}
\figcaption{
Comparison between the synthetic profiles calculated with the 1D model 
atmosphere
of Procyon (solid line) against the observations (filled circles). The residuals
(predicted -- observed) are also shown, but multiplied by a factor of 5.
\label{prd1}}
\end{center}
\end{figure*}

\begin{figure*}
\begin{center}
\includegraphics[width=12cm,angle=0]{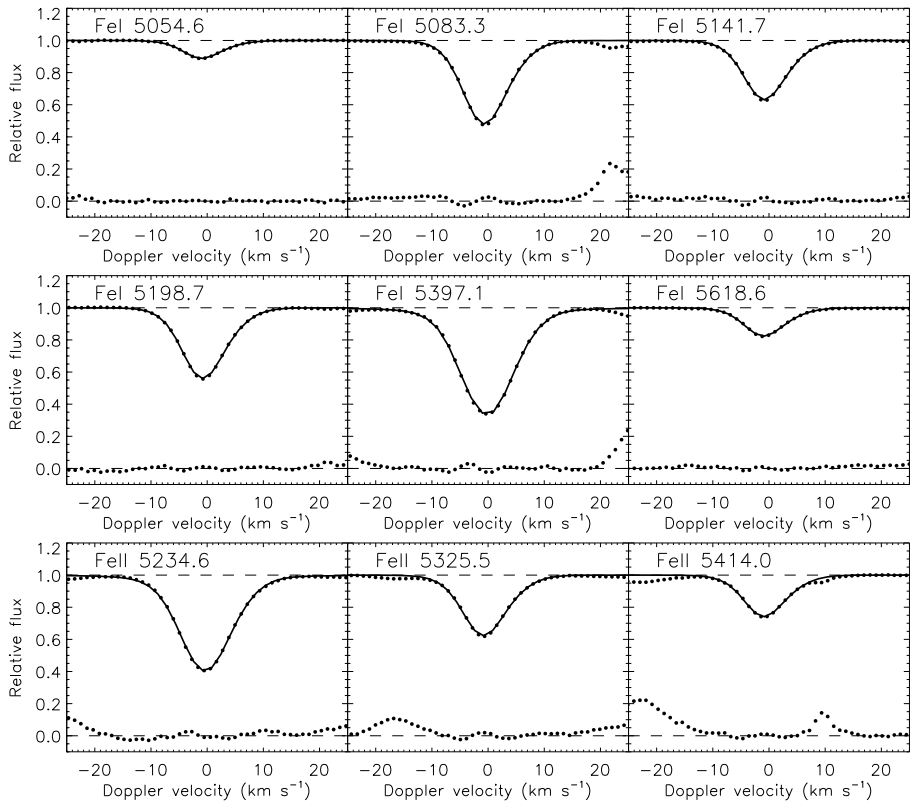}
\figcaption{
Comparison between the synthetic profiles calculated with the 3D model atmosphere
of Procyon against the observations  (filled circles). The residuals
(predicted -- observed) are also shown, but multiplied by a factor of 5.
\label{prd3}}
\end{center}
\end{figure*}

\begin{figure*}[t!]
\begin{center}
\includegraphics[width=9cm,angle=0]{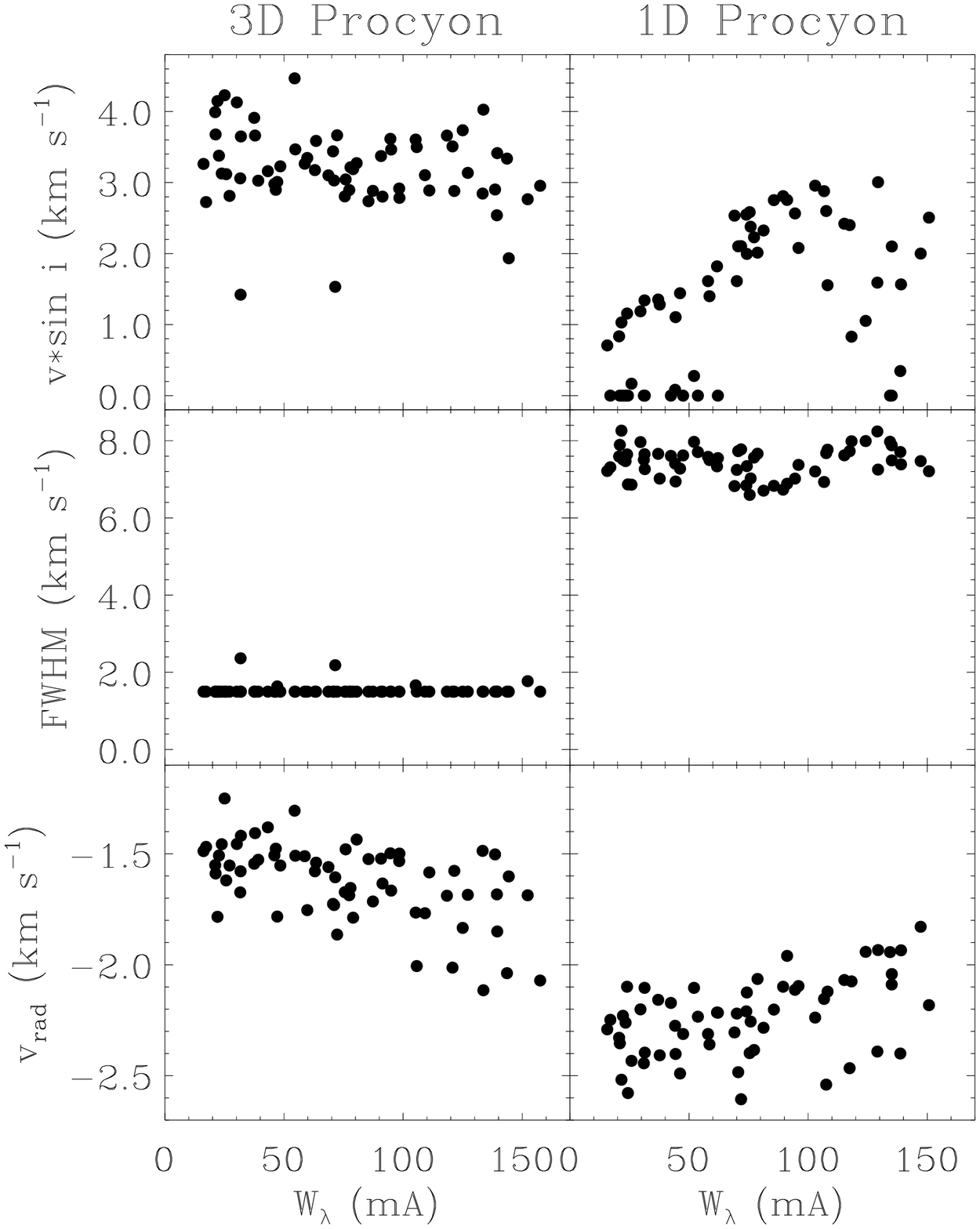}
\figcaption{
Optimal values for $v_{\rm rot} \sin i$, the 
FWHM of a Gaussian profile convolved to the 
synthetic line profile, and the velocity shift ($v_{\rm rad}$) to 
match the model and observed profiles of Procyon. The three panels on the
left correspond to the 3D model atmosphere, whereas the three on the
right to the 1D model assuming a microturbulence of 2.2 km s$^{-1}$.
\label{Fe_chi2_Procyon}}
\end{center}
\end{figure*}

In the case of 3D analyses, $v_{\rm rot}{\rm sin}i$ can
be determined by minimizing the residuals between theoretical and observed
profiles, since no other unknown broadening
like macro- and microturbulence enters the calculations. 
This is performed through a $\chi^2$-analysis with the elemental abundance,
$v_{\rm rot} \sin i$, FWHM, 
and $v_{\rm rad}$ as free parameters.  By 
FWHM we refer to that of 
an assumed-Gaussian\footnote{For the FTS solar observations
the instrumental profile is better described by a sinc function, but this has
a negligible impact on the calculated profiles.}
 profile that in 3D accounts for the
instrumental profile, 
while $v_{\rm rad}$ is the 
required overall velocity shift of the whole observed profile. 
In the similar 1D analysis described below the FWHM represents the combined
effect of instrumental resolution and macroturbulence,
 while the microturbulence
is kept fixed in the $\chi^2$-analysis. 
Figs. \ref{prd1} and \ref{prd3} compare the 
 observations and the best-fit
synthetic spectra calculated  with
1D and 3D model atmospheres, 
respectively, for several iron lines. 
A common deficiency of the 1D predictions is apparent from Fig. 8, where the
residuals show an oscillation around the line center. 
Fig. \ref{Fe_chi2_Procyon} shows the best-fit 
parameters obtained from  
individual lines.

For Procyon, we estimate 
$v_{\rm rot}{\rm sin}i = 3.16 \pm 0.50$\,km\,s$^{-1}$,
where the quoted uncertainty is the standard deviation 
for the clean samples of Fe I  and Fe II lines minus a few obviously
discrepant cases (see Fig. \ref{Fe_chi2_Procyon}).
There is, except for  a couple of aberrant lines, 
no need for additional Gaussian broadening above the known instrumental 
broadening 
($1.5$\,km\,s$^{-1}$); macroturbulence, which must be introduced with the 1D
model, is not required. 
The residual fluxes  show no 
systematic behavior
across the profile, as the corresponding 1D lines do. In fact, the 3D
modeling makes it possible to detect blended lines which would 
have escaped detection in 1D analyses.

\begin{figure*}[b!]
\begin{center}
\includegraphics[width=9cm,angle=0]{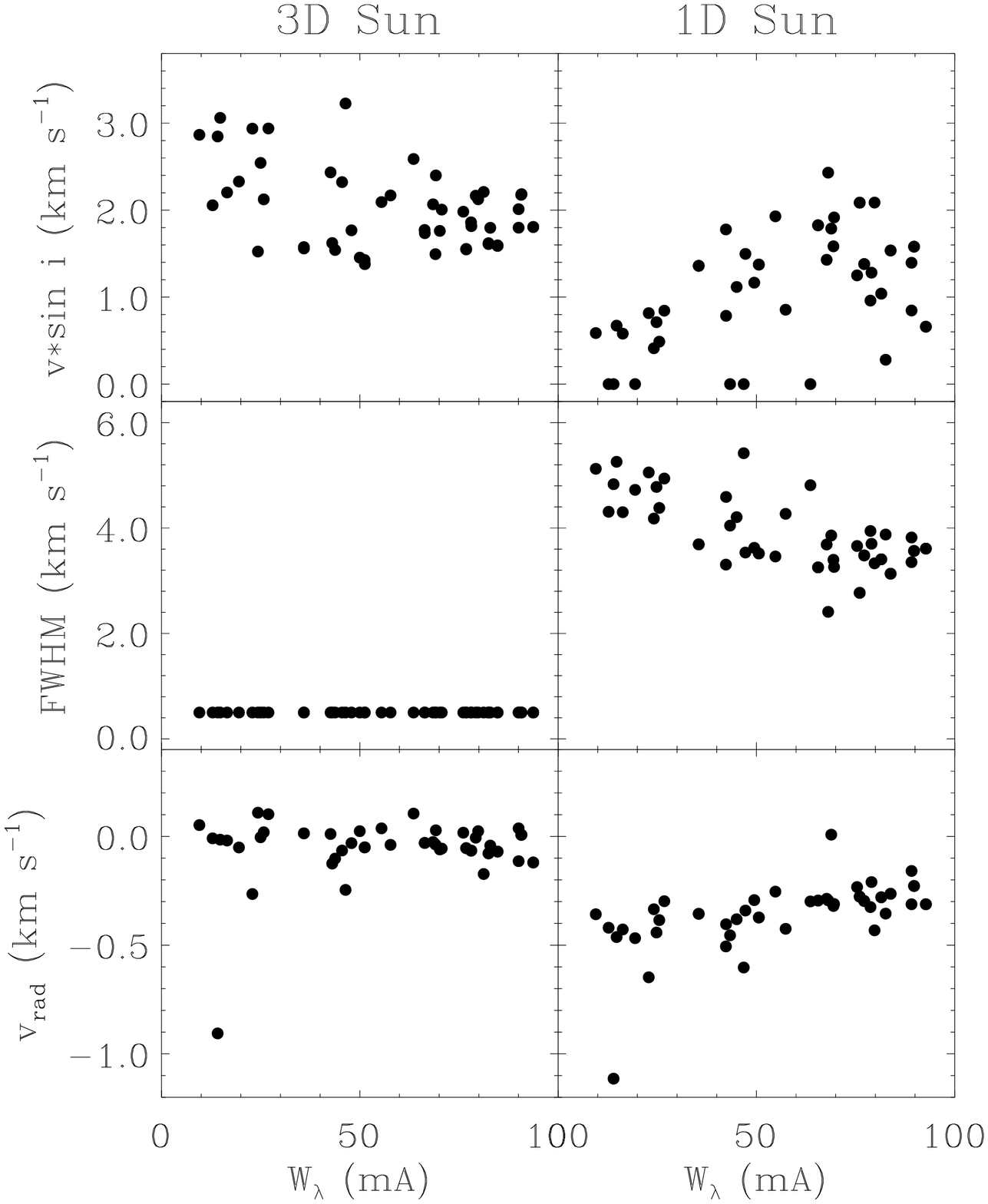}
\figcaption{
 Optimal values for $v_{\rm rot} \sin i$, the 
 FWHM of a Gaussian profile convolved to the 
 synthetic line profile, and the velocity shift ($v_{\rm rad}$) to 
 match the model and observed profiles of Sun. The three panels on the
 left correspond to the 3D model atmosphere, whereas the three on the
 right to the 1D model.
\label{Fe_chi2_Sun}}
\end{center}
\end{figure*}

It should be noted, however, that the above estimate of 
$v_{\rm rot}{\rm sin}i$
may be slightly over-estimated due to the finite numerical resolution of the
convection simulation. With  higher resolution more of the high-velocity
tails of the velocity distributions are sampled, which causes 
additional
broadening of the lines (Asplund et al. 2000a). For the Sun, the predicted
line profiles appear to have converged at the currently highest affordable
resolution ($200 \times 200 \times 82$). Since the present Procyon 
simulations are
based on a grid size of $100 \times 100 \times  82$ due to computing time
considerations, the widths of the theoretical line profiles may have been
slightly under-estimated. Such a conclusion is supported 
by the presence 
of a persistent
 trend in derived Fe abundances with line strength (Sect. \ref{abu}).
Assuming a similar behavior with numerical resolution
for Procyon as for the Sun, $v_{\rm rot}{\rm sin}i$ may be 
$\approx 0.5$\,km\,s$^{-1}$ less than estimated above. 
However, this additional broadening has a very limited impact 
on the predicted line shifts and asymmetries (Asplund et al. 2000a), in 
particular for  weak lines, which are still reliable indicators of the
chemical abundances.

A detailed Fourier analysis by Gray (1981b), using classical 1D model 
atmospheres and
McDonald spectra with a resolution of 2.5\,km\,s$^{-1}$, provided  
$v_{\rm rot}{\rm sin}i = 2.8 \pm 0.3$\,km\,s$^{-1}$, and a radial-tangential 
macroturbulence of $7.0 \pm 0.1$\,km\,s$^{-1}$.  
 Studies by Fekel (1997) and Benz \& Mayor (1984) gave estimates for 
 $v_{\rm rot}{\rm sin}i$ of $4.9 \pm 1.0$ and $4.5 \pm 1.1$\,km\,s$^{-1}$, 
 respectively.
 Our own $\chi^2$-analysis using 1D model atmospheres reveals similar 
 values but
also that the derived $v_{\rm rot}{\rm sin}i$ depends on the choice of 
line, in
particular for weak lines. At least partly, this will depend on the form of
the adopted macroturbulence, for example, Gaussian as here,
 or radial-tangential as used by Gray (1981b).
The scatter in $v_{\rm rot}{\rm sin}i$ is clearly significantly larger with 1D than with
3D model atmospheres when using otherwise identical analyses.
The reason why lines with $W_\lambda \simeq 70-100$\,m\AA\ apparently 
indicate the correct $v_{\rm rot}{\rm sin}i$, while weaker lines do not,
 remains a mystery to us; in fact the solar analysis shows the same effect 
 (Fig. \ref{Fe_chi2_Sun}).

\subsection{Line asymmetries and shifts}
\label{chifs}

A major advantage with using 3D hydrodynamical model atmospheres
for spectral synthesis besides the elimination of free parameters like
mixing length parameters, micro- and macroturbulence, is that the convective
velocity field and temperature inhomogeneities are self-consistently computed.
As a consequence, the imprint of convection in the form of asymmetries and
shifts of spectral lines can be predicted and compared with observations.
Furthermore, the predicted line asymmetries are on an
absolute wavelength scale. Since different lines have different 
sensitivity to the photospheric
structure and convective Doppler shifts, 
 all lines show unique asymmetries.

\begin{figure*}[t!]
\begin{center}
\includegraphics[width=7cm,angle=0]{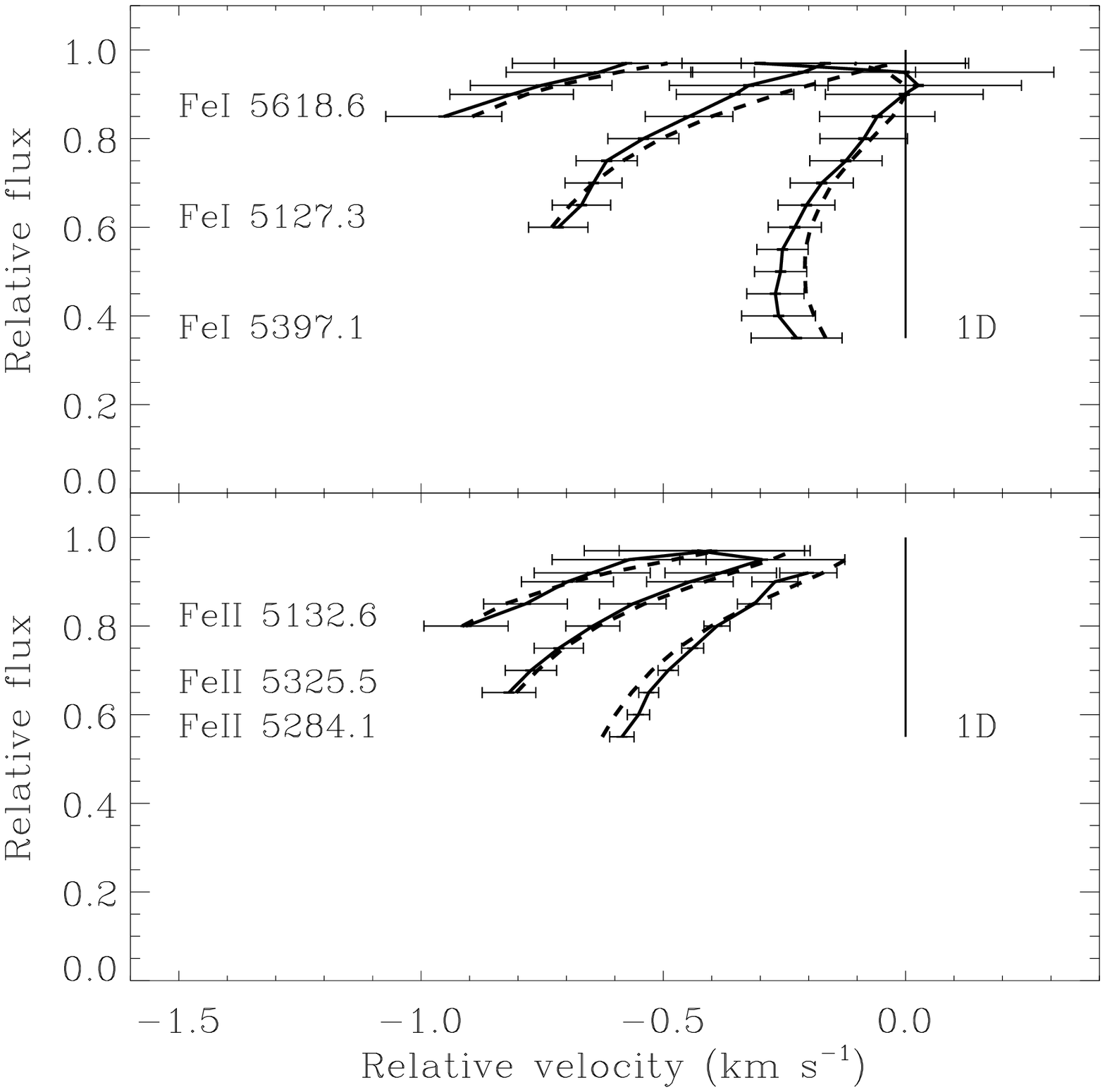}
\figcaption{
Comparison between line bisectors of three iron lines
 measured in the observed spectrum of Procyon 
(solid lines with error bars) with 
those measured in the synthetic spectrum calculated from the
3D model atmosphere (dashed lines).
\label{bisectors_ex}}
\end{center}
\end{figure*}

Figure \ref{bisectors_ex} illustrates the excellent 
agreement between predicted and observed line asymmetries for both Fe
I  and Fe II  lines and weak and strong lines.  Since the
exact radial velocity of Procyon is unknown, the observed spectrum is not
on an absolute wavelength scale. Therefore, the zero-point for all
observed bisectors in Fig. \ref{bisectors_ex} has been shifted by the same
amount 
to make them coincide with the theoretical
bisectors on average (see further discussion below and in Sect. 8). 
It should be noted, however, that the observed {\it relative}
shifts between different lines shown in Fig. \ref{bisectors_ex} have been
derived  
from the observations and are obviously well predicted by the
3D calculations.

Although the general agreement between theory and observations is quite 
satisfactory, there are several notable exceptions for which the
discrepant bisectors may signal problems like blends or an erroneous
laboratory wavelength. In the case of Fe I  4745.8\,\AA, Fe I
5434.5\,\AA, and Fe I  4576.3\,\AA\ the reason is most likely minor
blends in the wings, which is supported by the good correspondence
between predicted and observed bisectors for larger line depths.
The Fe II  4629.339\,\AA\ line can be suspected to have
an erroneous laboratory wavelength due to its uncharacteristically 
large offset of about 9\,m\AA\ to the red. 
It is noteworthy that this line is
the only Fe II  line in our sample which is not included in the
compilation of S. Johansson (1998, private communication) and 
the laboratory wavelength has instead been adopted from the VALD database.

\begin{figure*}
\begin{center}
\includegraphics[width=7cm,angle=0]{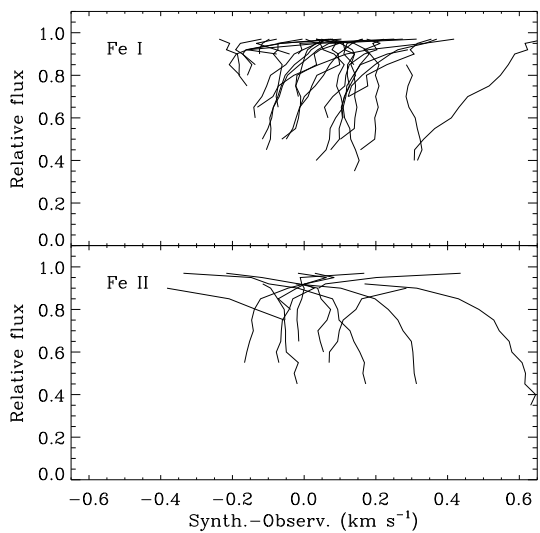}
\figcaption{
Difference between predicted and observed Fe I (upper panel) and
Fe II (lower panel) bisectors.
\label{Fe_dbis}}
\end{center}
\end{figure*}

The bisector comparison is summarized in Figure \ref{Fe_dbis}, 
which shows the 
{\it difference} between the theoretical and observed bisectors 
for the clean
samples of Fe I  and Fe II  lines together with a few additional 
unblended Fe II 
lines which lack an accurate determination of the transition probability 
and therefore
are not included in the estimate of the Fe abundance presented 
below (see \S \ref{abu}). 
For most lines, the predicted bisector is simply offset from the observed 
bisector, i.e.
the difference is essentially a vertical line, without  trend with line 
depth that could signal an improper theoretical velocity field. 
The scatter around zero difference is, at least partly, due to 
errors in the laboratory wavelengths, as well as 
the finite $S/N$  and spectral resolution of the observations.
Close to the continuum, the effects of weak blends become significant, 
which causes
a larger scatter and in a few cases a  deviating bisector. 
Typically, the {\it shapes} of the bisectors can be predicted to within  
50\,m\,s$^{-1}$,
while the typical velocity spans are 500\,m\,s$^{-1}$ or more,
which is quite remarkable considering the sensitivity of line asymmetries
upon the detailed atmospheric structure. In fact, the agreement is even slightly better
than for the Sun (Asplund et al. 2000b).

A particularly interesting result arising from the line asymmetry
comparison, is the predicted blueward hook close to the continuum in
the bisectors of some lines.    All
predicted Fe lines with $h < 0.5$, and hence partly
saturated, show this behavior, while it is not present in any weaker
lines.   The blue hook is induced  by the 
influence of extended Lorentzian wings from  the upflowing and hence
blue-shifted granules (Dravins 1990).  Due to the bias in line strength
towards granules as a result of  their steeper temperature
gradients, lines tend to first saturate and develop damping wings
there. The extended nature of these blue-shifted wings can thus
influence the resulting bisector shape, in particular close to the
continuum.  Even though blends tend to increase 
the uncertainty in the
measured bisectors close to the continuum,   there also exists 
observational evidence for this blue-ward hook. In
particular Fe I  5415.2\,\AA\ is a convincing case but also Fe I
4903.3\,\AA, Fe I  5397.1\,\AA,  and Fe II  4620.5\,\AA\ support this
conclusion. From the  
Procyon atlas of Griffin \& Griffin (1979), Dravins (1987) claimed to
have identified a similar feature by averaging over many lines,
however, the feature is not apparent in the average bisectors discussed
 in Sect. \ref{la}.

\begin{figure*}
\begin{center}
\includegraphics[width=7cm,angle=0]{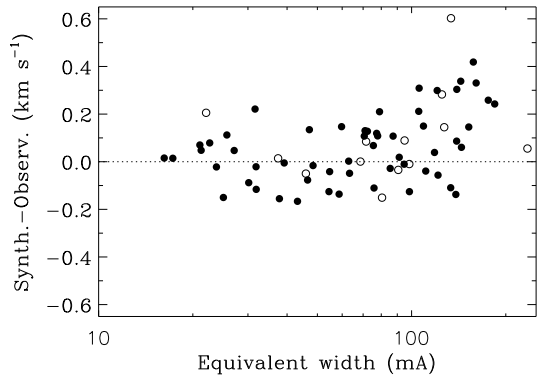}
\figcaption{ 
Difference between predicted and observed Fe I (filled circles) and
Fe II (open circles) line shifts. 
\label{Fe_dshift}} 
\end{center}
\end{figure*}

As expected from the good general agreement in terms of line 
asymmetries, the predicted differential line shifts 
match closely to the observations.
Fig. \ref{Fe_dshift} shows the difference between 
predicted and observed line shifts for
the Fe I  and Fe II  lines; the Fe I  4903.3101\,\AA\ and 
Fe II  4629.3390\,\AA\ lines stand out; probably with erroneous
laboratory wavelengths. Since the radial velocity of Procyon
is unknown, the observed profiles have  been 
shifted by $-1.56$\,km\,s$^{-1}$
in order to make the difference between 
predictions and observations disappear
on average. The mean of the differences between predicted and 
observed line shifts amounts 
to 0.00\,km\,s$^{-1}$ with $\sigma = 0.10$\,km\,s$^{-1}$,
implying that the radial velocity can be equally 
well determined in 3D analyses 
from fitting the whole profiles (Fig. \ref{Fe_chi2_Procyon}) or only 
the line center. The observed weakening of the convective 
blue-shift for stronger lines 
is the result of the the disappearance of the convective 
inhomogeneities as 
the line-formation
region is gradually moved outwards.
Unfortunately, the present observations  do not include
 sufficiently strong lines
to determine if  a plateau at zero convective 
line shifts exists, 
like that found for the Sun
(Allende Prieto \& Garc\'{\i}a L\'opez 1998), and the 
3D convection simulation can not be invoked either to
answer this question. As clear from Fig. \ref{Fe_dshift}, 
the predicted line shifts become systematically biased 
 for stronger lines, which is not surprising given the likely 
departures
from LTE in the cores of such lines. Besides, the 3D simulations lose realism 
for the upper atmosphere as a result of the influence of the upper boundary and 
missing magnetic fields. In those layers,  the presence of 
propagating waves and the fact that the energy balance  is determined
by a few strong lines out of LTE complicates quite seriously the modeling. 
A similar deficiency was detected in the solar study (Asplund et al. 2000b).

\begin{figure*}[t!]
\begin{center}
\includegraphics[width=7cm,angle=90]{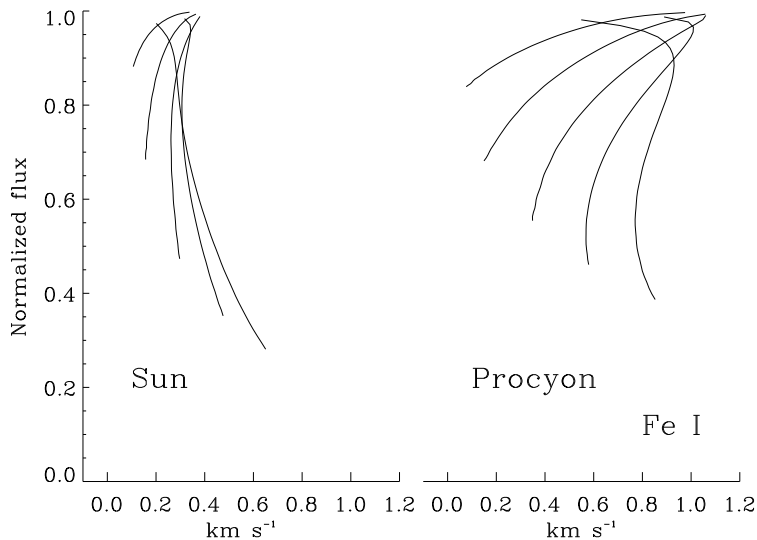}
\figcaption{  
Average bisectors for theoretical Fe I lines produced in the time-dependent
hydrodynamical 3D model atmosphere with a continuum normalized flux
at the line center f: $h< 0.35$, $0.35< h < 0.55$, $0.55< h < 0.75$, and
$h> 0.75$. To simulate lines of different strength, we adopted
a range of iron abundances in the calculation of a single line profile.
\label{bisectors_theory}} 
\end{center}
\end{figure*}

\begin{figure*}[ht!]
\begin{center}
\includegraphics[width=10cm,angle=90]{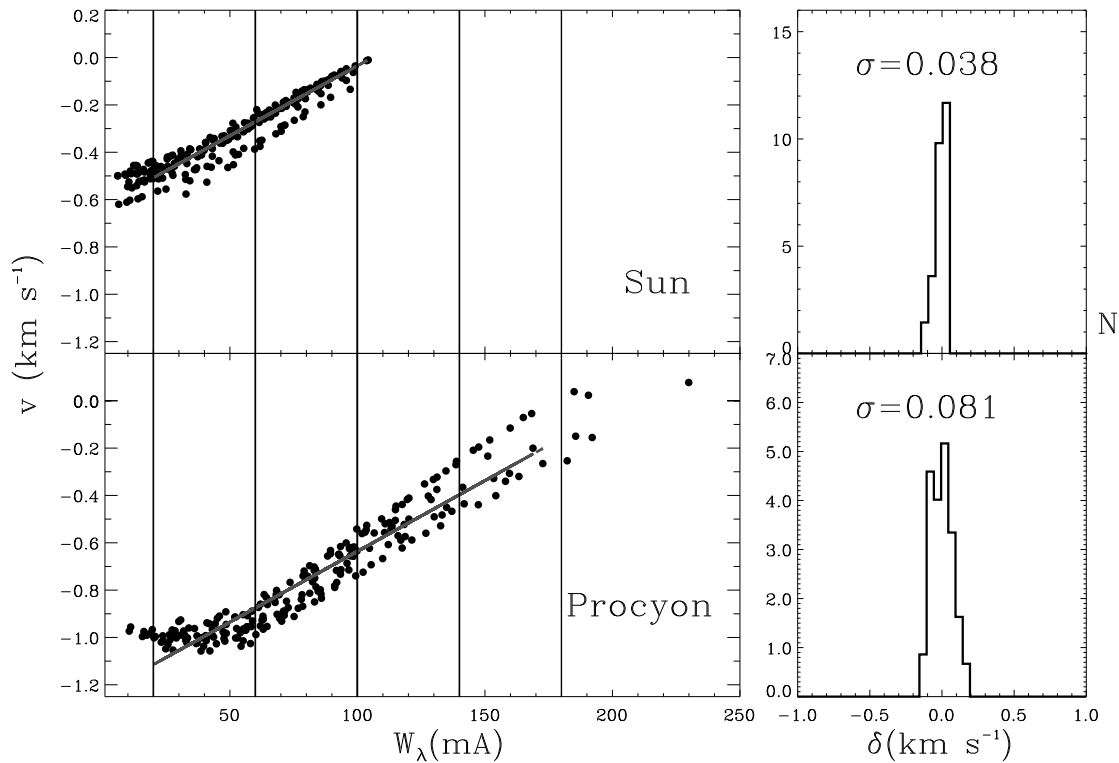}
\figcaption{  
Left panels: predicted velocity shifts between the stellar and 
laboratory wavelengths for relatively weak lines. Right panels: 
difference between the measured shifts and the linear fits to 
lines with $20 \le W_{\lambda} \le 180$ m\AA\ (histogram); the width
of the distribution is quantified by the $\sigma$ of a Gaussian
fit to the data.
\label{shifts_low_theory}} 
\end{center}
\end{figure*}

In Figure \ref{bisectors_theory}
we have mimicked Fe I lines of different strengths by altering the iron 
abundance in the calculations. Although a quantitative 
comparison is not possible, because of the different ways in which the two
plots were constructed, the general behavior 
of Fe I line asymmetries shown in Figure \ref{bisectors} 
is clearly reproduced by the theoretical calculations. 
We have calculated the 
line velocity shifts (v)
for a large group of Fe I lines, and as Figure \ref{shifts_low_theory} 
reveals,
the intrinsic scatter of the convective shifts for weak lines 
about a straight fit is significantly smaller in the Sun than in Procyon. This
result confirms qualitatively the observational findings in \S \ref{ls}. 
It should be noted, however, that since the same lines have not been used for
the theoretical and observed distributions one should not be surprised of
minor differences in the results.
 The predicted slopes of the velocity shifts as a function of equivalent width
 are $5.9  \times 10^{-3} $ for the Sun and 
 $6.0  \times 10^{-3}$ for Procyon -- virtually identical.
 Neither of them are very close to the observed slopes, 
 but the agreement is clearly better for Procyon, whose observed
 value is $4.9 \times 10^{-3}$, than for the Sun 
 ($3.2  \times 10^{-3}$; see Asplund et al. 2000a).

The line asymmetry comparison
clearly shows that the adopted 3D model atmosphere provides
quite a realistic description of the granulation properties on Procyon. 
We stress again that 1D models can predict neither the bisectors 
nor the shifts of lines. 

\section{The iron abundance of Procyon}
\label{abu}

The chemical abundances inferred from the comparison between 
our spectroscopic observations and the 3D hydrodynamical 
model atmosphere of Procyon have a special interest, as this is 
the most 
common driver for stellar spectroscopy. It is particularly 
useful to refer our
results to those from classical 1D model atmospheres, on which most
of the published stellar abundances are based. Here we determine the iron
abundance of Procyon 
with the two types of models (see \S \ref{ma}), making 
 a strictly differential comparison between them.

\subsection{Classical 1D models}
\label{1D}

\begin{figure*}[t!]
\begin{center}
\includegraphics[width=12cm,angle=90]{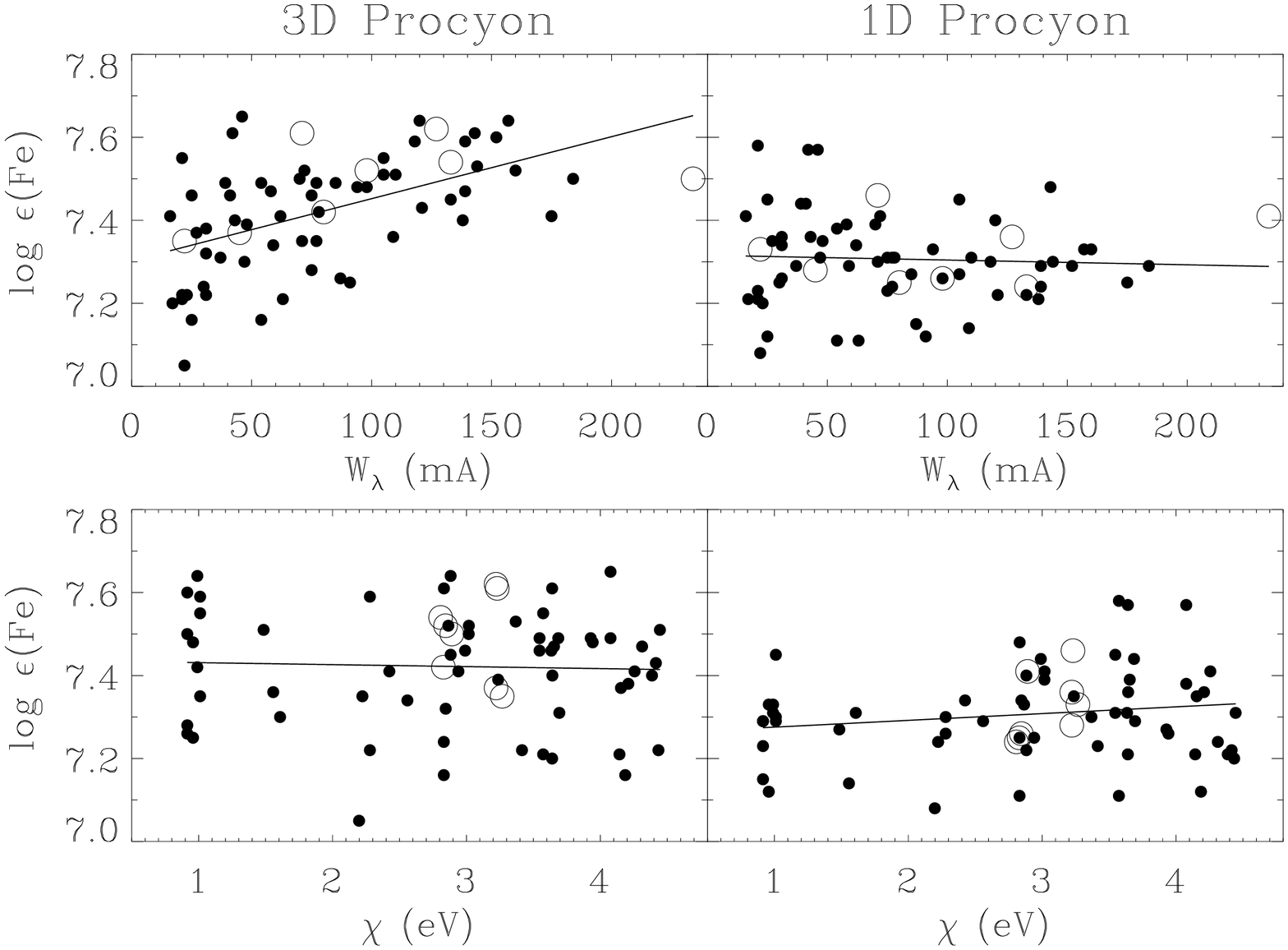}
\figcaption{  Iron abundance derived for Procyon 
from Fe I (filled circles) and Fe II lines
(open circles), as a function of the equivalent width (upper panel)
and the line excitation potential (lower panel).
\label{abu_procyon}} 
\end{center}
\end{figure*}

In addition to the free parameters in the $\chi^2$-analysis (iron abundance,
$v_{\rm rot} \sin i$, FWHM, and $v_{\rm rad}$), 
one must also specify the microturbulence ($\xi_{\rm tur}$) 
in analyses based on 1D model atmospheres. Variations in this parameter 
 affect both the line strengths
and shapes, in particular for partly saturated lines.
Following standard practice, we have  carried out several 
$\chi^2$-analyses
using different values of $\xi_{\rm tur}$ in order to select the value which 
removes any trend in
derived abundances with line strength. 
For Procyon, our choice of lines, transition probabilities, and model
 atmospheres 
suggest a value of $\xi_{\rm tur} = 2.2$\,km\,s$^{-1}$.
The results are displayed in the right-hand panels of Fig. \ref{abu_procyon}.
The mean abundance derived from the neutral-iron lines
 (filled circles) 
and ionized-iron lines (open circles) is $\log
 \epsilon$(Fe) $= 7.30 \pm 0.02$ ($\sigma = 0.11$) dex and $7.32 \pm 0.03$ 
 ($\sigma = 0.08$) dex, respectively. 

\begin{figure*}[b!]
\begin{center}
\includegraphics[width=12cm,angle=90]{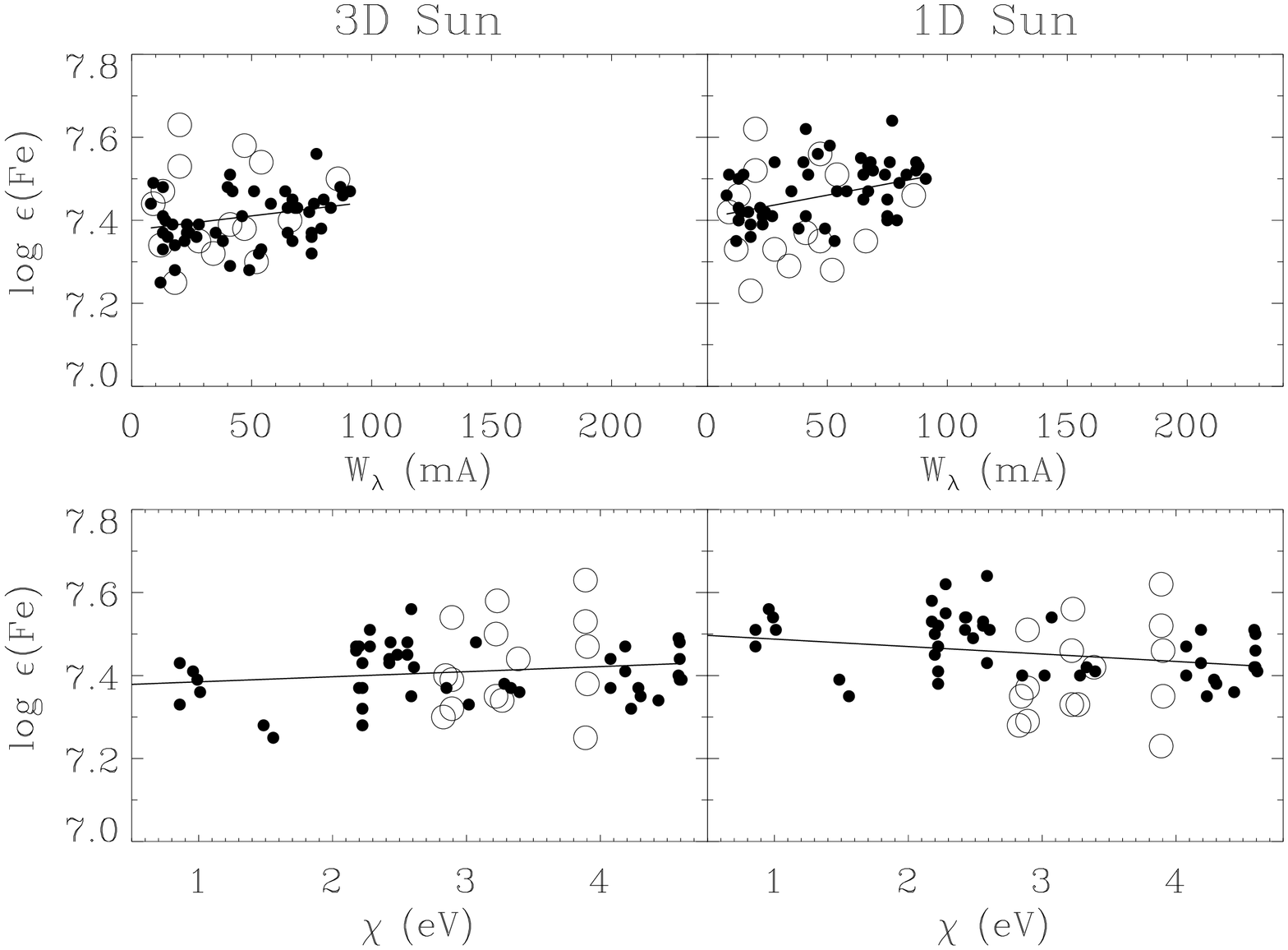}
\figcaption{
Iron abundance derived for the Sun 
from Fe I (filled circles) and Fe II lines
(open circles), as a function of the equivalent width (upper panel)
and the line excitation potential (lower panel).
\label{abu_sun}}
\end{center}
\end{figure*}

A similar analysis of the same lines used by Asplund et al. (2000c) in the
disk-integrated solar atlas (Kurucz et al. 1984), but 
 adopting the same sources for the radiative transition probabilities 
used for Procyon's (\S \ref{gfs}),
  provided  $\log \epsilon$ (Fe)$_{\odot} = 7.47  \pm  0.01 (\sigma = 0.07$) dex for the
 Fe I lines and $\log \epsilon$ (Fe)$_{\odot} = 7.41  \pm 0.03 
 (\sigma = 0.11$)
  dex for
 the Fe II lines with a {\sc marcs} model atmosphere and 
 $\xi = 1.0$ km s$^{-1}$. These figures are represented against equivalent width
 and excitation potential in the right-hand panels of Fig. \ref{abu_sun}.

\subsection{Three-dimensional models}
\label{3D}

Without an adjustable microturbulence parameter, the
process of determining the abundances from the line profiles
is straightforward from the $\chi^2$-analysis. 
To take advantage of all the information
 in the line profiles, and in parallel to our procedure for the 1D models, 
 we fit  $v_{\rm rot} \sin i$, FWHM,  and $v_{\rm rad}$, for 
 each individual line. Now, FWHM is also that of a Gaussian, but  it
  only accounts for the   instrumental profile,  
  as macro-turbulence is deemed as unnecessary (\S \ref{lpsr}). 
The left-hand panels of Figure \ref{abu_procyon} show
the  abundances
 from different lines plotted against the line equivalent width
and the line excitation potential. The mean abundance for
the neutral iron lines is $\log \epsilon$ (Fe)$_{\odot} = 7.41  \pm 0.02$
($\sigma = 0.14$) dex, and for the singly-ionized iron lines 
$\log \epsilon$ (Fe)$_{\odot} =
7.49  \pm 0.04$ ($\sigma = 0.10$) dex. The retrieved abundances tend to be 
higher for the stronger lines, as also found 
 for the
solar case with a more limited range of equivalent widths 
(see Fig. \ref{abu_sun}). 
This can be 
likely ascribed to the
limited resolution of the simulations, which miss the highest convective
velocities and therefore Doppler shifts. 
To put this into perspective we note that this trend is similar to an 
underestimated microturbulence of about 0.3 km s$^{-1}$ for Procyon, which
highlights that it is a relatively minor shortcoming of the 3D simulations
as most of the needed convective Doppler shifts are already accounted for.
Restricting the average to lines with equivalent widths
less or equal than 50 m\AA\ we obtain $7.35  \pm 0.03$ dex 
($\sigma=0.15$ dex; 22 lines) 
and $7.36 \pm 0.01$ dex ($\sigma=0.01$; 2 lines) for neutral 
and singly-ionized iron lines. 

 The analysis of the selected solar flux lines
previously used for the 1D case, 
 together 
with the $200 \times 200 \times 82$ 3D simulation described 
by Asplund et al. (2000b),  led to the following 
estimates of the solar iron abundance: 
$\log \epsilon$ (Fe)$_{\odot} = 7.40  \pm 0.01 $ ($\sigma = 0.07$) 
and $7.43 \pm $ 0.03 ($\sigma = 0.11$) dex,  
as derived 
from the neutral and singly-ionized iron lines, respectively. These figures
 change very mildly when we restrict the sample to lines with equivalent
widths smaller than 50 m\AA, as all the lines considered have equivalent widths
smaller than 100 m\AA.

The small differences in derived abundances with those given in 
Asplund et al. (2000c),
which are based on the same 3D model atmosphere as employed here, 
can be traced to slightly different 
adopted transition probabilities ($\la 0.02$ dex) and the use of flux profiles instead of
disk-center intensity profiles ($\la 0.02$ dex). Our new estimates do not 
supersede the solar iron abundances given in Asplund et al. (2000c). They  are   
  only intended to enable a differential study of Procyon.

\subsection{Discussion}

Estimates of the accuracy of the $\log gf$ for the transitions used are 
readily available, either from O'Brian et al. (1991), or from the tables in 
the Appendix. We can use them to perform an analysis similar to that in 
\S \ref{ls} for the line velocity shifts. This reveals that the errors in the
 $\log gf$s cannot account 
for more than about half of the scatter in the iron abundances. Even though
3D models need to be improved, there are probably other important  
error sources in our calculated line profiles. Departures from LTE are a likely 
candidate. We do not observe a larger scatter or a systematic deviation for
low-excitation Fe I lines that
one could expect from the 3D NLTE calculations of Trujillo Bueno \& Shchukina
(2001), but we notice a significantly larger scatter for the iron abundance
determined from Fe I lines in Procyon than in the Sun, and departures from
LTE are expected to grow for warmer stars.

Because of the shown imperfections in the 3D models, 
the preferred iron abundance for Procyon is that derived from the 3D 
analysis of weak lines: 
$\log \epsilon$ (Fe)$_{\odot} = 7.36  \pm 0.03 $ 
($\sigma = 0.15$). 

The main conclusions of our study of the iron abundance are:

\begin{enumerate}
\item The abundances derived from Fe I and Fe II lines are consistent both
for 1D and 3D analyses, and with a higher coherence for the 3D model. 
This implies that departures from LTE, which should be
minimal for Fe II, are likely small. Nevertheless, they could be responsible 
for a significant fraction of 
the scatter between the abundances retrieved from different lines, providing
a plausible explanation for the larger scatter observed 
for Procyon than for the Sun.

\item The differences between the iron abundances derived from 1D and 3D 
analyses (setting aside lines stronger than 50 m\AA\ in the case of Procyon)
are  small ($\lesssim 0.05$ dex), and the same conclusion applies to the solar case.
Other stars and elements may  show larger differences.

\item Procyon is marginally deficient in iron compared to the Sun by 
about 0.05 dex.

\end{enumerate}

\section{Center-to-limb variation}
\label{clvsect}

One of the applications of model atmospheres is to derive limb-darkening laws. 
These are required, for example, to correct interferometric measurements of
 stellar angular diameters, and affect directly the otherwise fully-empirical 
calibrations of effective temperature against color indices (see, e.g. 
Mozurkewich et al 1991).

\begin{figure*}
\begin{center}
\includegraphics[width=6cm,angle=0]{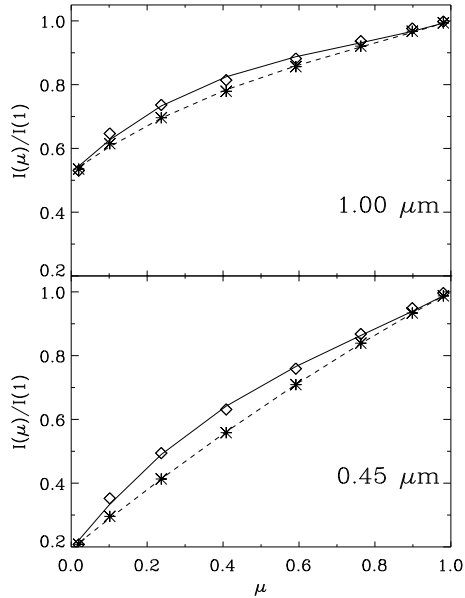}
\figcaption{
Limb darkening for the 3D (rhombi) and 1D (asterisk) models. Third-order
polynomials have been fitted by regular least squares to the data.
\label{clv}}
\end{center}
\end{figure*}

Here we compare the center-to-limb variation predicted by 
homogeneous models
 and the 3D simulations. Taking the emerging
 intensity for different angles at all different spatial locations we
produce a spatially and time-averaged intensity for different ray
inclinations.  Figure \ref{clv} shows the predicted limb-darkening for the
 1D and 3D models. Two wavelengths were selected, 4500
 and 10000 \AA. Strong differences are obvious in the plot, the
 limb-darkening being markedly non-linear for the inhomogeneous model.  We
 have fitted the data to third-order polynomials by regular least
 squares (see Equation 6 in Hanbury Brown et al. 1974) and the results
 correspond to the solid (3D) and dashed (1D) lines in Figure \ref{clv}. Lower 
 order polynomials, were inappropriate for the 3D model atmosphere. Hanbury
 Brown et al. (1974) calculated  the interferometric correlation
 factor  associated with this particular limb-darkening law.
At  4500 \AA, their determined angular diameter for a uniformly emitting
disk should be corrected by factors of 1.081 and 1.064 for the 1D and 3D cases,
 respectively. The 1D correction obtained from an older version of {\sc marcs}
  (Gustafsson et al. 1975) used by
 Mozurkewich et al. (1991)  was 1.07 at 4500 \AA.
The radius and the effective temperature of Procyon we derived in
 \S \ref{sp} should therefore be slightly corrected from 2.071 to 2.059
 $R_{\odot}$ and from 6512 to 6530 K, respectively. 
 The correction between 3D and 1D, which amounts to roughly 1.6 \%, implies 
 a correction to the effective temperature of roughly 
 50 K for a star like Procyon A, showing 
 the importance of detailed model atmospheres for establishing a 
 truly empirical $T_{\rm eff}$ scale from  measurements of 
 angular diameters (see, e.g., Di Benedetto 1998).

\section{Absolute Radial Velocity}

When discussing measurements of stellar radial velocities, it is
commonly assumed that an observed shift of the spectral features from
their  wavelengths at rest can be directly translated to a velocity
projected along the line of sight. Lindegren, Dravins \& Madsen (1999)
have recently discussed the dangers in doing so, especially when it
comes to precise measurements, pointing out effects that may need to be 
considered.

The gravitational shift is proportional to the gravitational field,
$V_g = G M/(Rc)$, and for late-type dwarfs
amounts to about 0.5 km s$^{-1}$. 
In the case of
Procyon A, we have a special advantage, as its mass, radius, and proper
motion are well determined. The gravitational redshift, including the
blueshift induced by  Earth is 
$V_g = 0.436 \pm 0.019$ km s$^{-1}$,
where we have used the corrected stellar radius 
$R= 2.06 \pm 0.02 R_{\odot}$ (see \S \ref{clvsect}). 
The transverse Doppler
effect scales with the square  of the modulus of the 
velocity,  $V_{tD} =
|V_t|^2/(2c)$, and becomes significant, reaching $\sim$ 0.25 km s$^{-1}$  for
the fastest halo stars orbiting at $\sim$ 400 km s$^{-1}$. 
The proper motion of Procyon as measured by {\it Hipparcos} 
corresponds to
a velocity perpendicular to the line of sight of $V_t = 20.87 \pm 0.07$ km s$^{-1}$,
which translates to a negligible {\it transverse} 
Doppler shift of $< 1$ m s$^{-1}$.

The list of contributors to the observed 
line shifts that have little  to do with the star's space 
motion is much
 longer: pulsations, stellar winds, surface convection, pressure
 gradients, etc.  In the case of {\it quiet} late-type dwarfs,  stellar
winds are not expected to
 disturb the photospheric line wavelengths, and the integrated light
 pulsations are known to have  periods  of minutes and amplitudes of a
 few meters per second or less, as detected in the Sun (see, e.g., Fossat \&
 Ricort 1975, Claverie et al. 1979) or Procyon itself (Marti{\' c} et al.
 1999).  Pressure gradients have been shown to produce line asymmetries
and shifts
 of only a few meters per second for solar-like stars (Allende Prieto,
 Garc\'{\i}a L\'opez \& Trujillo Bueno 1997). In conclusion, the main
 disturbing element is the  velocity pattern of convective origin.  The
detailed three-dimensional hydrodynamical simulations here presented
provide
 an accurate description of the convective velocity-temperature fields
 in the photosphere, and therefore constitute a means to zero the
 photospheric velocities.
  Confronting the measured line shifts with the predictions of 
  the 3D model (see  
  Section \ref{chifs}), and restricting the analysis 
  to lines weaker than 100 m\AA, 
  we determine 
 a velocity shift of the observed spectrum of Procyon A by $-1.56 \pm 0.13$
 km s$^{-1}$.  
That amount has to be divided between the  remaining
 effects expected to be important in the case of Procyon A:
 gravitational redshift, orbital velocity, and systemic velocity.

  Combining the astrometric orbital elements of Girard et al. (1996) 
 with the radial velocity amplitude measured by Irwin et al. (1992):
 $1.401 \pm 0.093$ km s$^{-1}$ (from absolute measurements)
 or $1.482 \pm 0.110$ km s$^{-1}$ (from relative determinations),
 we find that the orbital velocity of Procyon A  changed by less 
  than 0.0004 km s$^{-1}$ during the two days the observations 
  were carried out, and its value 
 was 1.115 (from absolute measurements) or 1.180 km s$^{-1}$ 
 (from relative determinations). Adopting  
 $1.15 \pm 0.10$ km s$^{-1}$, we obtain for the heliocentric 
 radial velocity
 of the system: $V_r = -1.56 (\pm 0.13) -0.44 (\pm 0.02) 
 - 1.15 (\pm 0.10) = -3.15 \pm 0.17$ km s$^{-1}$.

\section{Summary and Conclusions}

We have observed the spectrum of Procyon A (F5IV) 
from 4559 to 5780 \AA\ with a $S/N$ of $\sim 10^3$ and a 
resolving power of $2 \times 10^5$. We
have measured the line bisectors and relative line shifts 
of large number of Fe I and Fe II lines.  Compared to the Sun, 
the convective velocity shifts of the lines are significantly larger
and show a larger scatter from a linear relationship against equivalent
width.  The difference between the weakest and the strongest lines
 measured 
can reach up to 1 km s$^{-1}$, which is almost twice as large as the 
range observed in the Sun. Among other interesting consequences, 
this large differential velocities are likely to limit  the
 accuracy in the determination of absolute radial velocities in F-type
 stars.

We have constructed a new three-dimensional, time-dependent
hydrodynamical model atmosphere, and tested it against
the observed spectrum. The new model reproduces in detail most of the
 observed features, but we identify some room for improvement. 
At all levels, the comparison with the observed spectral lines shows a
significant improvement compared to classical homogeneous models. The
predicted line shifts and asymmetries are reasonably 
close to the observed ones. Our study of the iron abundance in the photosphere
of Procyon reveals that:
 
\begin{itemize}
\item The abundances derived from Fe I and Fe II lines are consistent both
for 1D and 3D analyses, and to a higher degree for the latter.
 This implies that departures from LTE, which should be
minimal for Fe II, are relatively small.

\item the differences between the mean iron abundances derived from 1D 
and 3D analysis are small ($\lesssim 0.05$ dex).

\item we find the iron abundance of Procyon to be 
$\log \epsilon ({\rm Fe}) = 7.36 \pm 0.03$ ($\sigma = 0.15$) dex,
concluding, from a differential study of the solar spectrum, 
that the star
is marginally deficient in iron compared to the Sun by $\sim 0.05$ dex.

\end{itemize}

We notice a scatter in the 
abundances from different lines that clearly exceeds the expectations from 
errors in the 
transition probabilities. Known imperfections in the
3D model should account for  part of the excess errors. Departures from
LTE are a potentially important source of error. 3D models with improved 
resolution and detailed
NLTE calculations are needed. In addition, other deficiencies in the input 
physics could contribute, and could be identified through meticulous 
tests against  observations.

The detailed
line shapes predicted by the model provide a reliable upper limit for
 the small projected rotational velocity of the star, 
$v_{\rm rot} \sin i \le 3.16$ km s$^{-1}$, with the correct value probably
close to 2.7 km s$^{-1}$.
The 3D model atmosphere shows a limb-darkening law that is 
noticeably different from the predictions of 1D models. This finding 
 has limited impact on the interferometric determination of the angular 
diameter of Procyon ($\sim 2$ \%) and its effective temperature 
($\sim$ 50 K), but is obviously important in establishing an absolute scale
of $T_{\rm eff}$ based on direct measurements. The ability of the 
hydrodynamical model atmosphere to
predict the convective shifts of the spectral lines, 
combined with the accurate parallax and proper motions
determined by {\it Hipparcos}, the published orbital elements, and 
 a careful assessment of the 
gravitational redshift  makes it possible to determine the 
space motion of the Procyon binary  system within 0.17  
km s$^{-1}$.

\acknowledgements

We are indebted to the staff at McDonald Observatory, in particular
David Doss, for invaluable assistance in the data acquisition. 
Sveneric Johansson and Ward  Whaling kindly delivered experimental 
Fe II $f$-values and line wavelengths, respectively. 
 \AA ke Nordlund, Robert Stein, and Regner Trampedach
are thanked for expert help with the 3D convection simulations. 

NSO/Kitt Peak FTS data used here were produced by NSF/NOAO. We have 
made use of the Hipparcos Catalogue, NASA's Astrophysics Data System 
Abstract Service, the SIMBAD database at the CDS, and
the Vienna Atomic Line Database (VALD). This research has been 
partially supported by the NSF (grant AST-0086321), the Robert A. Welch 
Foundation of Houston, Texas, the Swedish Natural Science
Foundation (grant NFR F990/1999), and the Royal Swedish Academy
of Sciences.

\newpage

\appendix{A Compilation  of Laboratory Absolute Transition Probabilities 
 for  Fe II lines with $\lambda > 300$ nm.}

The experimental technique most  used combines measured branching 
ratios  and lifetimes of the upper levels to derive
 transition probabilities. Measurements of branching ratios of 
 high quality have been published by Kroll \& Kock (1987), Whaling
(as included in the compilation by Fuhr, Martin \& Wiese 1988 -- also private communication), Heise \&
Kock (1990), and Pauls, Grevesse, \& Huber (1990). Level lifetimes have
been derived with accuracies in the range 0.02--0.14 ns by Biemont et
al. (1991), Guo et al. (1992), and Schnabel, Kock \& Holweger (1999) 
using   laser-induced fluorescence. Available lifetimes have been 
compiled and combined in Table \ref{table4}. When
more than one measurement was available, we  averaged them out.
The  variation among the lifetime of the levels
within a term is remarkably small (except maybe  for the $^4F$ term), 
 as expected if  LS-coupling holds. Lower accuracy measurements  that include  $^4D_{1/2}$ and $^4F_{3/2}$ levels (providing for them 
$2.9 \pm 0.2$ and $3.7 \pm 0.2$, respectively),  support the use of the term average  values for those cases ($2.95 \pm 0.03$ and $3.70 \pm 0.15$).
The  branching fractions measured by Kroll \& Kock, Whaling, Heise,
 \& Kock, and Pauls et al. were converted to transition 
 probabilities using different sources for the lifetimes
  of the involved upper levels. The third through 
sixth columns of Table \ref{table5}   list their published $\log (gf)$s, 
the quoted uncertainties, and the lifetimes used for the different 
transitions. We have re-scaled the individual $\log (gf)$s using 
the adopted lifetimes of Table \ref{table4}, to produce a weighted 
averaged when more than a single
measure  was available. When inconsistent 
$\log (gf)$s were derived from Kroll \& Kock and  
Heise \& Kock, the latter were preferred, following Schnabel et al. The finally adopted $\log (gf)$s and they formal uncertainties are displayed in the last column of Table \ref{table5}.


\clearpage







\clearpage


\tablenum{1}
\tablecolumns{7}
\tablewidth{0pc}
\begin{deluxetable}{ccccccc}
\scriptsize
\tablecaption{Spectral Lines Selected for the Iron  Abundance Analysis of Procyon   \label{table1}}
\tablehead{
\colhead{Species} & \colhead{Wavelength}  & \colhead{EP} & \colhead{$\log gf$} &   \colhead{$\log \epsilon$ 3D} & \colhead{$\log \epsilon$ 1D} &  \colhead{W$_{\lambda}$(3D)}\\
& \colhead{(\AA)}  & \colhead{(eV)} &   \colhead{(dex)} & \colhead{(dex)}  & \colhead{(dex)}  &  \colhead{(m\AA)} \\ }
\startdata
Fe I   &    4602.00 &    1.608 &    -3.13 &     7.30 &     7.31 &     47 \\
Fe I   &    4602.94 &    1.485 &    -2.21 &     7.51 &     7.27 &    105 \\
Fe I   &    4635.85 &    2.845 &    -2.36 &     7.32 &     7.34 &     31 \\
Fe I   &    4683.56 &    2.831 &    -2.32 &     7.24 &     7.25 &     30 \\
Fe I   &    4690.14 &    3.686 &    -1.64 &     7.49 &     7.44 &     39 \\
Fe I   &    4710.28 &    3.018 &    -1.61 &     7.52 &     7.41 &     72 \\
Fe I   &    4735.84 &    4.076 &    -1.33 &     7.65 &     7.57 &     46 \\
Fe I   &    4741.53 &    2.831 &    -1.76 &     7.16 &     7.11 &     54 \\
Fe I   &    4745.80 &    3.654 &    -1.27 &     7.47 &     7.39 &     58 \\
Fe I   &    4779.44 &    3.415 &    -2.02 &     7.22 &     7.23 &     21 \\
Fe I   &    4786.81 &    3.017 &    -1.61 &     7.50 &     7.39 &     70 \\
Fe I   &    4788.76 &    3.237 &    -1.76 &     7.39 &     7.35 &     48 \\
Fe I   &    4789.65 &    3.546 &    -0.96 &     7.49 &     7.31 &     77 \\
Fe I   &    4798.27 &    4.186 &    -1.17 &     7.16 &     7.12 &     25 \\
Fe I   &    4802.88 &    3.642 &    -1.51 &     7.40 &     7.36 &     43 \\
Fe I   &    4832.73 &    3.640 &    -1.73 &     7.61 &     7.57 &     42 \\
Fe I   &    4844.01 &    3.546 &    -2.05 &     7.46 &     7.45 &     25 \\
Fe I   &    4872.14 &    2.882 &    -0.57 &     7.45 &     7.22 &    133 \\
Fe I   &    4903.31 &    2.882 &    -0.93 &     7.64 &     7.40 &    120 \\
Fe I   &    4918.99 &    2.865 &    -0.34 &     7.52 &     7.33 &    160 \\
Fe I   &    4985.25 &    3.928 &    -0.56 &     7.49 &     7.27 &     85 \\
Fe I   &    4994.13 &    0.915 &    -2.97 &     7.26 &     7.15 &     87 \\
Fe I   &    5006.12 &    2.832 &    -0.62 &     7.61 &     7.48 &    143 \\
Fe I   &    5014.94 &    3.943 &    -0.30 &     7.48 &     7.26 &     98 \\
Fe I   &    5022.79 &    2.990 &    -2.20 &     7.46 &     7.44 &     41 \\
Fe I   &    5049.82 &    2.279 &    -1.36 &     7.59 &     7.30 &    118 \\
Fe I   &    5054.64 &    3.640 &    -1.92 &     7.20 &     7.21 &     17 \\
Fe I   &    5079.74 &    0.990 &    -3.24 &     7.42 &     7.31 &     78 \\
Fe I   &    5083.34 &    0.958 &    -2.84 &     7.25 &     7.12 &     91 \\
Fe I   &    5127.36 &    0.915 &    -3.25 &     7.28 &     7.23 &     75 \\
Fe I   &    5141.74 &    2.424 &    -2.24 &     7.41 &     7.34 &     62 \\
Fe I   &    5145.09 &    2.198 &    -2.88 &     7.05 &     7.08 &     22 \\
Fe I   &    5151.91 &    1.011 &    -3.32 &     7.35 &     7.30 &     71 \\
Fe I   &    5194.94 &    1.557 &    -2.02 &     7.36 &     7.14 &    109 \\
Fe I   &    5198.71 &    2.223 &    -2.09 &     7.35 &     7.24 &     77 \\
Fe I   &    5232.94 &    2.940 &    -0.06 &     7.41 &     7.25 &    175 \\
Fe I   &    5242.49 &    3.634 &    -0.97 &     7.46 &     7.31 &     75 \\
Fe I   &    5288.52 &    3.694 &    -1.51 &     7.31 &     7.29 &     37 \\
Fe I   &    5321.11 &    4.434 &    -1.09 &     7.22 &     7.20 &     23 \\
Fe I   &    5322.04 &    2.279 &    -2.80 &     7.22 &     7.26 &     31 \\
Fe I   &    5328.04 &    0.915 &    -1.47 &     7.50 &     7.29 &    184 \\
\tablebreak
Fe I   &    5364.87 &    4.445 &     0.23 &     7.51 &     7.31 &    110 \\
Fe I   &    5365.40 &    3.573 &    -1.02 &     7.21 &     7.11 &     63 \\
Fe I   &    5367.47 &    4.415 &     0.44 &     7.43 &     7.22 &    121 \\
Fe I   &    5383.37 &    4.312 &     0.64 &     7.47 &     7.24 &    139 \\
Fe I   &    5397.13 &    0.915 &    -1.98 &     7.60 &     7.29 &    152 \\
Fe I   &    5403.82 &    4.076 &    -1.03 &     7.49 &     7.38 &     54 \\
Fe I   &    5415.20 &    4.386 &     0.64 &     7.40 &     7.21 &    138 \\
Fe I   &    5434.52 &    1.011 &    -2.13 &     7.59 &     7.29 &    139 \\
Fe I   &    5446.92 &    0.990 &    -1.91 &     7.64 &     7.33 &    157 \\
Fe I   &    5464.28 &    4.143 &    -1.40 &     7.21 &     7.21 &     21 \\
Fe I   &    5466.99 &    3.573 &    -2.23 &     7.55 &     7.58 &     21 \\
Fe I   &    5483.10 &    4.154 &    -1.41 &     7.37 &     7.35 &     27 \\
Fe I   &    5497.52 &    1.011 &    -2.83 &     7.55 &     7.45 &    105 \\
Fe I   &    5501.47 &    0.958 &    -3.05 &     7.48 &     7.33 &     94 \\
Fe I   &    5586.76 &    3.368 &    -0.14 &     7.53 &     7.30 &    144 \\
Fe I   &    5618.63 &    4.209 &    -1.27 &     7.38 &     7.36 &     31 \\
Fe I   &    5701.54 &    2.559 &    -2.14 &     7.34 &     7.29 &     59 \\
Fe I   &    5741.85 &    4.256 &    -1.67 &     7.41 &     7.41 &     16 \\
Fe II  &    4576.33 &    2.844 &    -2.91 &     7.52 &     7.26 &     98 \\
Fe II  &    4620.51 &    2.828 &    -3.19 &     7.42 &     7.25 &     80 \\
Fe II  &    4629.34 &    2.807 &    -2.28 &     7.54 &     7.24 &    133 \\
Fe II  &    4923.92 &    2.891 &    -1.26 &     7.50 &     7.41 &    234 \\
Fe II  &    5234.62 &    3.221 &    -2.23 &     7.62 &     7.36 &    127 \\
Fe II  &    5264.80 &    3.230 &    -3.23 &     7.61 &     7.46 &     71 \\
Fe II  &    5414.07 &    3.221 &    -3.48 &     7.37 &     7.28 &     45 \\
Fe II  &    5525.12 &    3.267 &    -3.94 &     7.35 &     7.33 &     22 \\
\enddata
\end{deluxetable}

\clearpage

\tablenum{2}
\tablecolumns{5}
\tablewidth{0pc}
\begin{deluxetable}{ccccccc}
\scriptsize
\tablecaption{Spectral Lines Selected for the Iron  Abundance Analysis of the Sun   \label{table2}}
\tablehead{
\colhead{Species} & \colhead{Wavelength}  & \colhead{EP} & \colhead{$\log gf$} &   \colhead{$\log \epsilon$ 3D} & \colhead{$\log \epsilon$ 1D} &  \colhead{W$_{\lambda}$(3D)}\\
& \colhead{(\AA)}  & \colhead{(eV)} &   \colhead{(dex)} & \colhead{(dex)}  & \colhead{(dex)}  &  \colhead{(m\AA)} \\ }
\startdata
 Fe I  &    4389.25 &    0.052 &    -4.55 &     7.36 &     7.45 &     75 \\
 Fe I  &    4445.47 &    0.087 &    -5.38 &     7.29 &     7.41 &     41 \\
 Fe I  &    5044.21 &    2.851 &    -2.02 &     7.37 &     7.40 &     75 \\
 Fe I  &    5247.05 &    0.087 &    -4.98 &     7.43 &     7.54 &     68 \\
 Fe I  &    5250.21 &    0.121 &    -4.90 &     7.35 &     7.47 &     67 \\
 Fe I  &    5253.46 &    3.283 &    -1.57 &     7.38 &     7.40 &     79 \\
 Fe I  &    5329.99 &    4.076 &    -1.22 &     7.44 &     7.47 &     58 \\
 Fe I  &    5412.79 &    4.434 &    -1.72 &     7.34 &     7.36 &     18 \\
 Fe I  &    5491.83 &    4.186 &    -2.19 &     7.41 &     7.43 &     13 \\
 Fe I  &    5525.54 &    4.230 &    -1.08 &     7.32 &     7.35 &     53 \\
 Fe I  &    5661.35 &    4.284 &    -1.76 &     7.37 &     7.39 &     23 \\
 Fe I  &    5701.54 &    2.559 &    -2.14 &     7.48 &     7.52 &     87 \\
 Fe I  &    5705.46 &    4.301 &    -1.36 &     7.35 &     7.38 &     38 \\
 Fe I  &    5778.45 &    2.588 &    -3.44 &     7.35 &     7.43 &     22 \\
 Fe I  &    5784.66 &    3.396 &    -2.53 &     7.36 &     7.41 &     27 \\
 Fe I  &    5855.08 &    4.607 &    -1.48 &     7.39 &     7.41 &     23 \\
 Fe I  &    5956.69 &    0.859 &    -4.50 &     7.33 &     7.47 &     54 \\
 Fe I  &    6082.71 &    2.223 &    -3.55 &     7.37 &     7.47 &     35 \\
 Fe I  &    6136.99 &    2.198 &    -2.93 &     7.37 &     7.45 &     65 \\
 Fe I  &    6151.62 &    2.176 &    -3.37 &     7.47 &     7.58 &     51 \\
 Fe I  &    6173.34 &    2.223 &    -2.88 &     7.43 &     7.52 &     69 \\
 Fe I  &    6200.31 &    2.608 &    -2.37 &     7.42 &     7.51 &     74 \\
 Fe I  &    6219.28 &    2.198 &    -2.45 &     7.47 &     7.50 &     91 \\
 Fe I  &    6240.65 &    2.223 &    -3.17 &     7.28 &     7.38 &     49 \\
 Fe I  &    6265.13 &    2.176 &    -2.54 &     7.46 &     7.53 &     88 \\
 Fe I  &    6271.28 &    3.332 &    -2.70 &     7.37 &     7.42 &     24 \\
 Fe I  &    6280.62 &    0.859 &    -4.39 &     7.43 &     7.51 &     65 \\
 Fe I  &    6297.79 &    2.223 &    -2.64 &     7.32 &     7.41 &     75 \\
 Fe I  &    6322.69 &    2.588 &    -2.47 &     7.56 &     7.64 &     77 \\
 Fe I  &    6481.87 &    2.279 &    -3.01 &     7.47 &     7.55 &     64 \\
 Fe I  &    6498.94 &    0.958 &    -4.69 &     7.41 &     7.56 &     46 \\
 Fe I  &    6574.23 &    0.990 &    -5.02 &     7.39 &     7.54 &     28 \\
 Fe I  &    6581.21 &    1.485 &    -4.68 &     7.28 &     7.39 &     18 \\
 Fe I  &    6593.87 &    2.433 &    -2.37 &     7.48 &     7.54 &     87 \\
 Fe I  &    6609.11 &    2.559 &    -2.66 &     7.45 &     7.53 &     67 \\
 Fe I  &    6625.02 &    1.011 &    -5.34 &     7.36 &     7.51 &     15 \\
 Fe I  &    6667.71 &    4.584 &    -2.11 &     7.49 &     7.51 &      9 \\
 Fe I  &    6699.14 &    4.593 &    -2.10 &     7.44 &     7.46 &      8 \\
 Fe I  &    6739.52 &    1.557 &    -4.79 &     7.25 &     7.35 &     12 \\
 Fe I  &    6750.15 &    2.424 &    -2.59 &     7.44 &     7.54 &     76 \\
 Fe I  &    6793.26 &    4.076 &    -2.33 &     7.37 &     7.40 &     13 \\
 Fe I  &    6804.27 &    4.584 &    -1.81 &     7.40 &     7.42 &     14 \\
 Fe I  &    6837.01 &    4.593 &    -1.69 &     7.39 &     7.42 &     17 \\
 Fe I  &    6854.82 &    4.593 &    -1.93 &     7.48 &     7.50 &     13 \\
 Fe I  &    6945.21 &    2.424 &    -2.45 &     7.43 &     7.51 &     83 \\
 Fe I  &    6971.93 &    3.018 &    -3.34 &     7.33 &     7.40 &     13 \\
 Fe I  &    6978.85 &    2.484 &    -2.45 &     7.45 &     7.49 &     80 \\
 Fe I  &    7189.15 &    3.071 &    -2.77 &     7.48 &     7.54 &     40 \\
 Fe I  &    7401.69 &    4.186 &    -1.60 &     7.47 &     7.51 &     42 \\
 Fe I  &    7723.21 &    2.279 &    -3.62 &     7.51 &     7.62 &     41 \\
 \tablebreak
Fe II  &    4576.33 &    2.844 &    -2.91 &     7.40 &     7.35 &     66 \\
Fe II  &    4620.51 &    2.828 &    -3.19 &     7.30 &     7.28 &     52 \\
Fe II  &    4656.98 &    2.891 &    -3.58 &     7.32 &     7.29 &     34 \\
Fe II  &    5234.62 &    3.221 &    -2.23 &     7.50 &     7.46 &     86 \\
Fe II  &    5264.80 &    3.230 &    -3.23 &     7.58 &     7.56 &     47 \\
Fe II  &    5414.07 &    3.221 &    -3.48 &     7.35 &     7.33 &     28 \\
Fe II  &    5525.12 &    3.267 &    -3.94 &     7.34 &     7.33 &     12 \\
Fe II  &    5627.49 &    3.387 &    -4.10 &     7.44 &     7.42 &      9 \\
Fe II  &    6432.68 &    2.891 &    -3.51 &     7.39 &     7.37 &     41 \\
Fe II  &    6516.08 &    2.891 &    -3.38 &     7.54 &     7.51 &     54 \\
Fe II  &    7222.39 &    3.889 &    -3.37 &     7.63 &     7.62 &     20 \\
Fe II  &    7224.48 &    3.889 &    -3.29 &     7.53 &     7.52 &     20 \\
Fe II  &    7449.33 &    3.889 &    -3.07 &     7.25 &     7.23 &     18 \\
Fe II  &    7515.83 &    3.903 &    -3.45 &     7.47 &     7.46 &     13 \\
Fe II  &    7711.72 &    3.903 &    -2.45 &     7.38 &     7.35 &     47 \\
\enddata
\end{deluxetable}

\clearpage

\tablenum{3}
\tablecolumns{7}
\tablewidth{0pc}
\begin{deluxetable}{lllllll}
\scriptsize
\tablecaption{Radiative Lifetimes for the Different Levels of Fe II \label{table4}}
\tablehead{
\colhead{Term}  & \colhead{J} & \colhead{Biemont et al} & \colhead{Guo et al } & \colhead{Schnabel et al} & \colhead{Adopted} & \colhead{Term average} \\
\colhead{}  & \colhead{} & \multicolumn{5}{c}{(ns)} }
\startdata
$^6D$&	9/2&	3.70 0.06&	      &		   &	3.70 0.06  &			\\
  &	7/2&	3.68 0.07&	      & 3.64 0.09   &	3.66 0.07   &			\\
  &	5/2&	3.63 0.08&	      &	3.70 0.05   &	3.68 0.05   &			\\
  &	3/2& 	3.83 0.10&	      &	3.73 0.07   &	3.76 0.07   &			\\
  &	1/2&	3.76 0.10&	      &		   &	3.76 0.10   &	3.71 $\sigma$=0.05	\\
\tableline
$^6F$&	11/2&	3.19 0.04&	      &		   &	3.19 0.04   &			\\
  &	9/2&	3.24 0.06&	      &		   &	3.24 0.06   &			\\
  & 	7/2&	3.26 0.10&	      &		   &	3.26 0.10   &			\\
  &	5/2&	3.33 0.09&	      &		   &	3.33 0.09   &			\\
  &	3/2& 	3.34 0.10&	      &		   &	3.34 0.10   &	3.27 $\sigma$=0.06 \\
\tableline
$^6P$&	7/2&	3.73 0.05& 3.73 0.06 &		   &	3.73 0.05   &			\\
  &	5/2&	3.83 0.07& 3.79 0.12 &		   &	3.81 0.07   &			\\
  &	3/2&		 & 3.71 0.12 &		   &	3.71 0.12   &	3.75 $\sigma$=0.05	\\
\tableline
$^4P$&	5/2&		 & 3.43 0.09 &		   &	3.43 0.09   &			\\
  &	3/2&		 & 3.44 0.11 &		   &	3.44 0.11   &	3.44 $\sigma$=0.01 \\
\tableline				
$^4D$&	7/2&		 & 3.02 0.07 &	2.97 0.02   &	2.97 0.02   &			\\
  &	5/2&		 & 3.10 0.08 &	2.90 0.06   &	2.97 0.06   &			\\
  &	3/2&		 &	     &	2.91 0.09   &	2.91 0.09   &			 \\
  &     1/2&		 &           &		    &	2.95 0.03\tablenotemark{a}& 2.95 $\sigma$=0.03\\
\tableline
$^4F$&	9/2&		 & 3.87 0.09 &		    &	3.87 0.09   &			\\
  &     7/2&		 &  3.63 0.11&		    &	3.63 0.11   &			\\
  &     5/2&		 & 3.75 0.14 &	3.55 0.08   &	3.60 0.08   &			 \\
  &     3/2&		 &	     &		    &	3.70 0.15\tablenotemark{a}&3.70 $\sigma$=0.15  \\
\enddata
\tablenotetext{a}{Measurement not available. The term average is adopted. Note that  
Hannaford et al give $2.9 \pm 0.2$ ($^4$D) and $3.7 \pm 0.2$ ($^4$F)}
\end{deluxetable}

\clearpage

\tablenum{4}
\tablecolumns{7}
\tablewidth{0pc}
\begin{deluxetable}{ccccccc}
\scriptsize
\tablecaption{Oscillator Strengths for Fe II Lines in the Optical Region\label{table5}}
\tablehead{
\colhead{Wavelength}  & \colhead{Term} & \colhead{Kroll \& Kock} & \colhead{Whaling} & \colhead{Heise \& Kock} & \colhead{Pauls et al} & \colhead{Adopted} \\
\colhead{(\AA)}  & \colhead{} & \multicolumn{5}{c}{(dex dex ns)} }
\startdata
3002.64&  $^4P_{5/2}$& -0.93 0.07 3.9&		&		&		&	-0.87 0.07 \\
3163.09&  $^4F_{5/2}$& -2.82 0.07 4.0& 		&-2.82 0.07 4.0\tablenotemark{a}	&	&	-2.78 0.07 \\
3170.34&  $^4D_{1/2}$&		&		&	 	&-2.49 0.08 3.4	&	-2.42 0.08 \\
3183.11&  $^4F_{5/2}$& -2.10 0.07 4.0&-2.04 0.11  4.0&   		&		&	-2.04 0.06 \\
3185.31&  $^4F_{3/2}$& -2.78 0.07 4.1&		&		&		&	-2.74 0.07 \\
3186.74&  $^4D_{3/2}$& -1.67 0.04 3.2&-1.71 0.11  3.4 & 		&		&	-1.63 0.04 \\
3192.92&  $^4D_{5/2}$& -1.95 0.11 3.2&	   	&		&-1.92 0.07 3.4 & 	-1.88 0.06 \\
3193.80&  $^4D_{1/2}$&		&		&	 	&-1.75 0.07 3.4	&	-1.69 0.07 \\
3196.08&  $^4F_{7/2}$& -1.73 0.11 3.9&-1.66 0.11  3.9	& 		&		&	-1.67 0.08 \\
3210.45&  $^4D_{3/2}$& -1.69 0.04 3.2&-1.79 0.11  3.4 &-1.69 0.04 3.2\tablenotemark{a}&	&	-1.66 0.04 \\
3213.31&  $^4D_{5/2}$& -1.27 0.07 3.2&	   	&		& -1.31 0.07 3.4& 	-1.25 0.05 \\
3227.73&  $^4D_{7/2}$& -1.06 0.11 3.7&-1.13 0.04  3.7 & 		&		&	-1.02 0.04 \\
3255.89&  $^6D_{7/2}$& -2.52 0.04 4.2&-2.50 0.22  4.0&-2.52 0.04 4.2\tablenotemark{a}	&	&	-2.46 0.04 \\
3277.35&  $^6D_{9/2}$& -2.30 0.04 3.9&-2.47 0.11  3.9	&		&		&	-2.31 0.04 \\
322.774&  $^4D_{7/2}$&	       &-1.13 0.07  3.7	&		&		&	-1.03 0.07 \\
3281.30&  $^6D_{5/2}$&-2.69 0.04 4.1& 		&-2.69 0.04 4.1\tablenotemark{a}	&	&	-2.65 0.04 \\
3285.41&  $^6D_{1/2}$&-2.87 0.04 3.9&-2.12 0.06  4.0	&		&		&	-2.82 0.04 \\
3295.81&  $^6D_{3/2}$&-2.90 0.04 4.0&   	   	&-2.90 0.04 4.0\tablenotemark{a}	&	&	-2.87 0.04 \\
3302.86&  $^6D_{7/2}$&-3.51 0.07 4.2&            	&		&		&	-3.45 0.07 \\
3303.47&  $^6D_{1/2}$&-2.70 0.11 3.9&			&		&		&	-2.68 0.11 \\
4173.47&  $^4D_{5/2}$&-2.18 0.07 3.2&		   	&	 	&-2.77 0.14 3.4	&	-2.50 0.08 \\
4178.87&  $^4F_{7/2}$&-2.48 0.11 3.9&		&		&		&	-2.45 0.11 \\
4233.17&  $^4D_{7/2}$&-1.91 0.07 3.7&-2.00 0.11  3.7	&-1.91 0.07 3.7\tablenotemark{a}	&	&	-1.84 0.06 \\
4303.17&  $^4D_{3/2}$&-2.49 0.07 3.2&		   	&-2.65 0.07 3.2	&		&	-2.61 0.07 \\
4351.76& $^4D_{5/2}$&-2.10 0.07 3.2&			&-2.10 0.07 3.2\tablenotemark{a}& -1.99 0.08 3.4	&-2.03 0.05 \\
4385.39&  $^4D_{1/2}$&		&		&	 	&-2.75 0.06 3.4	&	-2.69 0.06 \\
4491.40&  $^4F_{3/2}$&-2.70 0.11 4.1&			&		&		&	-2.66 0.11 \\
4508.29&  $^4D_{1/2}$&		&		&	 	&-2.58 0.09 3.4	&	-2.52 0.09 \\
4515.33&  $^4F_{5/2}$&-2.41 0.11 4.0	&	   	&		&		&	-2.36 0.11 \\
4522.62&  $^4D_{3/2}$&-2.03 0.11 3.2	&	   	&		&		&	-1.99 0.11 \\
4549.46&  $^4D_{5/2}$&-1.75 0.07 3.2	&	   	&		&-2.33 0.30 3.4	&	-1.87 0.08 \\
4555.88&  $^4F_{7/2}$&-2.29 0.11 3.9	&		&		&		&	-2.26 0.11 \\
4576.33&  $^4D_{5/2}$&          	&		&-2.94 0.09 3.2	&-2.97 0.16 3.4	&	-2.91 0.07 \\
4582.83&  $^4F_{7/2}$&-3.10 0.11 3.9&			&		&		&	-3.07 0.11 \\
4583.83&  $^4D_{7/2}$&-1.84 0.07 3.7&	-2.02 0.22  3.7 & 		&		&	-1.77 0.07 \\
4620.51&  $^4D_{7/2}$&              &			&-3.29 0.08 3.7	&	    	&	-3.19 0.07 \\
4629.34&  $^4F_{9/2}$&-2.33 0.11 4.3&			&		&		&	-2.28 0.11 \\
4656.97&  $^4D_{5/2}$&              & 		&-3.61 0.10 3.2	&	    	&	-3.58 0.08 \\
4923.92&  $^6P_{3/2}$&-1.24 0.11 4.0& -1.32 0.11  4.0	&		&		&	-1.26 0.08 \\
5169.03&  $^6P_{7/2}$&-0.87 0.11 3.8&			&		&		&	-0.86 0.11 \\
5197.57&  $^4F_{3/2}$&-2.10 0.11 4.1&			&		&		&	-2.06 0.11 \\
5234.66& $^4F_{5/2}$&-2.05 0.11 4.0&			&-2.27 0.10 4.0	\tablenotemark{d}&	&	-2.23 0.08 \\
5262.47&  $^4D_{5/2}$&              &     	   	&-3.06 0.10 3.2\tablenotemark{d}	&	&	-3.03 0.10 \\
5264.79&  $^4D_{3/2}$&          	&		&-3.27 0.08 3.2 &	    	&	-3.23 0.05 \\
\tablebreak
5276.00&  $^4F_{7/2}$&-1.94 0.11 3.9&			&		&		&	-1.91 0.11 \\
5316.62&  $^4F_{9/2}$&-1.85 0.11 4.3&			&		&		&	-1.85 0.11 \\
5316.78&  $^4D_{5/2}$&		&	   	&		&-2.80 0.12 3.4	&	-2.74 0.12 \\
5325.55&  $^4F_{7/2}$&-2.60 0.11 3.9&		&		&		&-2.57 0.11\tablenotemark{b} \\
5414.08&  $^4D_{7/2}$&          	&		&-3.58 0.09 3.7	&	    	&	-3.48 0.08 \\
5525.13&  $^4D_{7/2}$&          	&		&-4.04 0.11 3.7	&	    	&	-3.94 0.09 \\
5607.12&  $^6D_{7/2}$&		&     	   	&-3.83 0.07 4.2 &		&	-3.77 0.07 \\
5627.49&  $^4F_{5/2}$&       	   	&		&-4.14 0.09 4.0	&		&	-4.10 0.07 \\
6369.42&  $^6D_{3/2}$&		 &    	   	&-3.50 0.12 4.0	&	&	-3.47 0.12 \\
6432.67&  $^6D_{5/2}$&    	 	&		&-3.55 0.08 4.1	&	    	&	-3.51 0.06 \\
6516.07&  $^6D_{7/2}$&      	   	&		&-3.44 0.07 4.2	&	    	&	-3.38 0.05 \\
7222.39&  $^4D_{1/2}$&		&		&		&-3.43 0.07 3.4	&	-3.37 0.07 \\
7224.49&  $^4D_{1/2}$&		&		&		&-3.35 0.06 3.4	&	-3.29 0.06 \\
7449.34&  $^4D_{5/2}$&       	   	&		&-3.10 0.11 3.2	&	    	&	-3.07 0.09 \\
7515.83&  $^4D_{5/2}$&      	    	&		&-3.41 0.12 3.2& -3.53 0.08 3.4	&-3.45 0.07 \\
7711.73&  $^4D_{7/2}$&      	    	&		&-2.55 0.08 3.7	&	    	&	-2.45 0.07 \\
\enddata
\tablenotetext{a}{HK lines taken as reference from KK}
\tablenotetext{b}{This line has $\log gf = -3.2$ as computed by Kurucz. Only with a value similar to that we can reconcile the iron abundance from this line with that from the other Fe II lines}
 \tablenotetext{c}{Comparison with the solar spectrum shows that a value closer to -3.38 is to be preferred}
 \tablenotetext{d}{Line classified as a possible blend or a misidentification by Heise \& Kock}
\end{deluxetable}


\begin{thebibliography}{}

\bibitem[]{} Allende Prieto, C., Barklem, P. S., Asplund, M., \& Ruiz Cobo, B.
2001, \apj, 558, 830

\bibitem[Allende Prieto \& Garc\'{\i}a L\'opez(1998)]{AllendePrietoGarciaLopez98} Allende Prieto, 
C. \& Garc\'{\i}a Lopez, R. J.  1998, \aaps, 131, 431 

\bibitem[Allende Prieto, Garc\'{\i}a L\'opez, Lambert, \& 
Gustafsson(1999)]{1999ApJ...526..991A} Allende Prieto, C., Garcia Lopez, 
R.~J., Lambert, D.~L., \& Gustafsson, B.\ 1999a, \apj, 526, 991 

\bibitem[Allende Prieto et al.(1999b)]{AllendePrieto_et_al99b} Allende Prieto, 
C., Garc\'{\i}a Lopez, R. J., Lambert, D. L. \& Gustafsson, B. 1999b, \apj, 527, 879 


\bibitem[Allende Prieto \& Lambert(1999)]{1999A&A...352..555A} Allende 
Prieto, C.~\& Lambert, D.~L.\ 1999, \aap, 352, 555 

\bibitem[Allende Prieto, Garcia Lopez, \& Trujillo 
Bueno(1997)]{1997ApJ...483..941A} Allende Prieto, C., Garc\'{\i}a L\'opez, R.~J., 
\& Trujillo Bueno, J.\ 1997, \apj, 483, 941 

\bibitem[Anstee \& O'Mara(1991)]{1991MNRAS.253..549A} Anstee, S.~D.~\& 
O'Mara, B.~J.\ 1991, \mnras, 253, 549 

\bibitem{} Asplund M., 2000, A\&A 359, 755

\bibitem{} Asplund, M., \& Garc\'{\i}a P{\'e}rez, A.E. 2001, A\&A, 372, 601

\bibitem{} Asplund M., Gustafsson B., Kiselman D., Eriksson K., 1997, 
A\&A 318, 521

\bibitem{}  Asplund M., Nordlund \AA., Trampedach R., Stein R.F., 
1999, A\&A 346, L17

\bibitem{}  Asplund M., Ludwig H.-G., Nordlund \AA., Stein R.F., 
2000a, A\&A 359, 669

\bibitem{}  Asplund M., Nordlund \AA., Trampedach R., 
Allende Prieto C., Stein R.F., 2000b, A\&A 359, 729

\bibitem{}  Asplund M., Nordlund \AA., Trampedach R., Stein R.F., 
2000c, A\&A 359, 743

\bibitem[Atroshchenko \& Gadun(1994)]{1994A&A...291..635A} Atroshchenko, 
I.~N.~\& Gadun, A.~S.\ 1994, \aap, 291, 635 

\bibitem[Barklem, Piskunov, \& O'Mara(2000)]{2000A&A...363.1091B} Barklem, 
P.~S., Piskunov, N., \& O'Mara, B.~J.\ 2000, \aap, 363, 1091 

\bibitem[Barklem, O'Mara, \& Ross(1998)]{1998MNRAS.296.1057B} Barklem, 
P.~S., O'Mara, B.~J., \& Ross, J.~E.\ 1998, \mnras, 296, 1057 

\bibitem[Barklem \& O'Mara(1997)]{1997MNRAS.290..102B} Barklem, P.~S.~\& 
O'Mara, B.~J.\ 1997, \mnras, 290, 102 


\bibitem[Benz \& Mayor(1984)]{1984A&A...138..183B} Benz, W.\ \& Mayor, M.\ 
1984, \aap, 138, 183 

\bibitem[Bertelli et al.(1994)]{1994A&AS..106..275B} Bertelli, G., Bressan, 
A., Chiosi, C., Fagotto, F., \& Nasi, E.\ 1994, \aaps, 106, 275 

\bibitem[Biemont et al.(1991)]{1991A&A...249..539B} Biemont, E., Baudoux, 
M., Kurucz, R.~L., Ansbacher, W., \& Pinnington, E.~H.\ 1991, \aap, 249, 539 


\bibitem[Blackwell et al.(1986)]{1986MNRAS.220..549B} Blackwell, D.~E., 
Booth, A.~J., Haddock, D.~J., Petford, A.~D., \& Leggett, S.~K.\ 1986, 
\mnras, 220, 549 


\bibitem[Bruls \& Rutten(1992)]{1992A&A...265..257B} Bruls, J.~H.~M.~J.~\& 
Rutten, R.~J.\ 1992, \aap, 265, 257 

\bibitem[Claverie et al.(1979)]{1979Natur.282..591C} Claverie, A., Isaak, 
G.~R., McLeod, C.~P., van der Raay, H.~B., \& Cortes, T.~R.\ 1979, \nat, 
282, 591 


\bibitem[Chollet \& Sinceac(1999)]{1999A&AS..139..219C} Chollet, F.~\& 
Sinceac, V.\ 1999, \aaps, 139, 219 

\bibitem[Dravins(1987)]{Dravins87} Dravins, D. 1987, \aap, 172, 211 

\bibitem[Dravins(1990)]{1990A&A...228..218D} Dravins, D.\ 1990, \aap, 228, 
218 

\bibitem[]{} Dravins, D. 1999, in Precise Stellar Radial Velocities, J. B. Hearnshaw and C. D. Scarfe, eds., ASP Conf. Ser. No. 185, p. 268


\bibitem[Dravins Larsson \& Nordlund(1986)]{DravinsLarssonNordlund86} Dravins, 
D., Larsson, B. \& Nordlund, \AA. 1986, \aap, 158, 83 

\bibitem[Dravins Lindegren \& Nordlund(1981)]{DravinsLindegrenNordlund81} Dravins, 
D., Lindegren, L. \& Nordlund, \AA. 1981, \aap, 96, 345 

\bibitem[Dravins \& Nordlund(1990)]{1990A&A...228..184D} Dravins, D.~\& 
Nordlund, \AA.\ 1990, \aap, 228, 184 


\bibitem[]{} Edvardsson, B., Andersen, J., Gustafsson, B., Lambert, D. L., Nissen, P. E., \& Tomkin, J. 1993, \aap, 275, 101

\bibitem[Fekel(1997)]{1997PASP..109..514F} Fekel, F.\ C.\ 1997, \pasp, 109, 
514 

\bibitem[Fossat \& Ricort(1975)]{1975A&A....43..243F} Fossat, E.~\& Ricort, 
G.\ 1975, \aap, 43, 243 

\bibitem[Fuhr, Martin, \& Wiese(1988)]{1988atpi.book.....F} Fuhr, J.~R., 
Martin, G.~A., \& Wiese, W.~L.\ 1988, New York: American Institute of 
Physics (AIP) and American Chemical Society, 1988,  


\bibitem[Fuhrmann et al.(1997)]{Fuhrmann_et_al97} Fuhrmann, K., Pfeiffer, 
M., Frank, C., Reetz, J. \& Gehren, T. 1997, \aap, 323, 909 

\bibitem[Girard et al.(2000)]{2000AJ....119.2428G} Girard, T.~M.~et al.\ 
2000, \aj, 119, 2428 

\bibitem[Giridhar \& Ferro(1995)]{1995RMxAA..31...23G} Giridhar, S.~\& 
Ferro, A.~A.\ 1995, Revista Mexicana de Astronomia y Astrofisica, 31, 23 

\bibitem[Gray(1981)]{Gray81} Gray, D. F. 1981a, \apj, 251, 583 

\bibitem[Gray(1981)]{1981ApJ...251..152G} Gray, D.~F.\ 1981b, \apj, 251, 152 

\bibitem[Gray(1982)]{1982ApJ...255..200G} Gray, D.~F.\ 1982, \apj, 255, 200 

\bibitem[Gray \& Nagel(1989)]{GrayNagel89} Gray, D. F. \& Nagel, T.  
1989, \apj, 341, 421 


\bibitem[Griffin \& Griffin(1979)]{GriffinGriffin79} Griffin, R. \& 
Griffin, R. 1979, Cambridge: Institute of Astronomy, Observatories

\bibitem[Guo et al.(1992)]{1992PhRvA..46..641G} Guo, B., Ansbacher, W., 
Pinnington, E.~H., Ji, Q., \& Berends, R.~W.\ 1992, \pra, 46, 641 

\bibitem[Gustafsson, Bell, Eriksson, \& 
Nordlund(1975)]{1975A&A....42..407G} Gustafsson, B., Bell, R.~A., Eriksson, 
K., \& Nordlund, \AA.\ 1975, \aap, 42, 407 


\bibitem[Hamilton \& Lester(1999)]{1999PASP..111.1132H} Hamilton, D.~\& 
Lester, J.~B.\ 1999, \pasp, 111, 1132 

\bibitem[Hanbury Brown, Davis, \& Allen(1974)]{1974MNRAS.167..121H} Hanbury 
Brown, R., Davis, J., \& Allen, L.~R.\ 1974, \mnras, 167, 121 


\bibitem[Hauschildt Allard \& Baron(1999)]{HauschildtAllardBaron99} Hauschildt, 
P. H., Allard, F.  \& Baron, E. 1999, \apj, 512, 377 

\bibitem[Holweger \& Mueller(1974)]{1974SoPh...39...19H} Holweger, H.~\& 
M\"uller, E.~A.\ 1974, \solphys, 39, 19 

\bibitem[Kiselman \& Nordlund(1995)]{1995A&A...302..578K} Kiselman, D.~\& 
Nordlund, \AA.\ 1995, \aap, 302, 578 

\bibitem[Kroll \& Kock(1987)]{1987A&AS...67..225K} Kroll, S.~\& Kock, M.\ 
1987, \aaps, 67, 225 

\bibitem[]{} Kulander, J. L., \& Jefferies, J. T. 1966, \apj, 146, 194

\bibitem[Kupka et al.(1999)]{1999A&AS..138..119K} Kupka, F., Piskunov, N., 
Ryabchikova, T.~A., Stempels, H.~C., \& Weiss, W.~W.\ 1999, \aaps, 138, 119 

\bibitem[Kurucz(1992)]{Kurucz92} Kurucz, R. L. 1992, private communication

\bibitem[Kurucz Furenlid \& Brault (1984)]{KuruczFurenlidBrault84} Kurucz, R. 
L., Furenlid, I.  \& Brault, J.  1984, National Solar Observatory Atlas, 
Sunspot, New Mexico: National Solar Observatory, 1984,  

\bibitem[]{} Lambert, D. L., Heath, J. E., Lemke, M., \& Drake, J. 1006, 
\apjs, 103, 183

\bibitem[Lindegren, Dravins, \& Madsen(1999)]{1999psrv.conf...73L} 
Lindegren, L., Dravins, D., \& Madsen, S.\ 1999, ASP Conf.~Ser.~185: IAU 
Colloq.~170: Precise Stellar Radial Velocities, 73 

\bibitem[Marti{\' c} et al.(1999)]{1999A&A...351..993M} Marti{\' c}, M.~et 
al.\ 1999, \aap, 351, 993 

\bibitem[Mashonkina Shimanskaya \& Shimansky(1995)]{MashonkinaShimanskayaShimansky95} 
Mashonkina, L. I., Shimanskaya, N. N. \& Shimansky, V. V. 1995, ASP Conf. 
Ser. 78: Astrophysical Applications of Powerful New Databases, p. 389 

\bibitem[Mihalas, Dappen, \& Hummer(1988)]{1988ApJ...331..815M} Mihalas, 
D., Dappen, W., \& Hummer, D.~G.\ 1988, \apj, 331, 815 


\bibitem[Mozurkewich et al.(1991)]{1991AJ....101.2207M} Mozurkewich, D.~et 
al.\ 1991, \aj, 101, 2207 


\bibitem[Nave et al.(1994)]{Nave_et_al94} Nave, G., Johansson, S., 
Learner, R. C. M., Thorne, A. P. \& Brault, J. W. 1994, \apjs, 94, 221

\bibitem[Nordlund(1982)]{1982A&A...107....1N} Nordlund, \AA.\ 1982, \aap, 
107, 1 
 
 \bibitem[Nordlund \& Dravins(1990)]{1990A&A...228..155N} Nordlund, \AA.~\& 
Dravins, D.\ 1990, \aap, 228, 155 

\bibitem[O'Brian et al.(1991)]{1991OSAJB...8.1185O} O'Brian, T.~R., 
Wickliffe, M.~E., Lawler, J.~E., Whaling, W., \& Brault, J.~W.\ 1991, 
Optical Society of America Journal B Optical Physics, 8, 1185 

\bibitem[Pauls, Grevesse, \& Huber(1990)]{1990A&A...231..536P} Pauls, U., 
Grevesse, N., \& Huber, M.~C.~E.\ 1990, \aap, 231, 536 


\bibitem[Provencal et al.(1997)]{1997ApJ...480..777P} Provencal, J.~L., 
Shipman, H.~L., Wesemael, F., Bergeron, P., Bond, H.~E., Liebert, J., \& 
Sion, E.~M.\ 1997, \apj, 480, 777 


\bibitem[Rice \& Wehlau(1984)]{RiceWehlau84} Rice, J. B. \& Wehlau, 
W. H. 1984, \apj, 278, 721 


\bibitem[]{} Grevesse, N.~\& 
Sauval, A.~J.\ 1998, Space Science Reviews, 85, 161 


\bibitem[Schnabel, Kock, \& Holweger(1999)]{1999A&A...342..610S} Schnabel, 
R., Kock, M., \& Holweger, H.\ 1999, \aap, 342, 610 

\bibitem[Shchukina \& Trujillo Bueno(2001)]{2001ApJ...550..970S} Shchukina, 
N.~\& Trujillo Bueno, J.\ 2001, \apj, 550, 970 

\bibitem[Sneden(1973)]{Sneden73} Sneden, C. 1973, PhD Thesis, University of Texas at Austin


\bibitem[Stein \& Nordlund(1998)]{SteinNordlund98} Stein, R. F. \& 
Nordlund, \AA. 1998, \apj, 499, 914 


\bibitem[Thevenin(1989)]{Thevenin89} Th\'evenin, F. 1989, \aaps, 77, 
137 

\bibitem[Thevenin(1990)]{Thevenin90} Th\'evenin, F. 1990, \aaps, 82, 
179 

\bibitem[]{} Trampedach, R. 1997, Master Thesis, University of Aarhus

\bibitem[Tull et al.(1995)]{Tull_et_al95} Tull, 
R. G., MacQueen, P. J., Sneden, C.  \& Lambert, D. L. 1995, \pasp, 107, 251 

\bibitem[Unsold(1955)]{1955QB461.U55......} Uns\"old, A.\ 1955, Physik der
Sternatmosph\"aren, Berlin: Springer

\bibitem[Watanabe \& Steenbock(1985)]{WatanabeSteenbock85} Watanabe, T. \& 
Steenbock, W. 1985, \aap, 149, 21 


\end{thebibliography}
\end{document}